\documentclass[prx,a4paper,twocolumn,showpacs,floatfix,superscriptaddress,amsmath,amsfonts,amssymb,preprintnumbers,citeautoscript,aps,longbibliography]{revtex4-2}
\pdfoutput=1
\usepackage[T1]{fontenc}\usepackage[latin1]{inputenc}
\usepackage{dcolumn,graphicx,color,booktabs,microtype,afterpage} \graphicspath{{Figures/}}
\usepackage[charter,greekuppercase=italicized]{mathdesign}\usepackage{sidecap}
\renewcommand{\figurename}{Fig.}
\renewcommand{\tablename}{Table}

\newcount\hh \newcount\mm
\hh=\time \divide\hh by 60
\mm=\hh \multiply\mm by 60 \mm=-\mm
\advance\mm by \time
\def\now{\number\hh:\ifnum\mm<10{}0\fi\number\mm}

\usepackage{hyperref}
\hypersetup{colorlinks,plainpages=false,linkcolor=blue,urlcolor=blue,citecolor=blue,pdfpagemode=UseNone,pdfstartview=FitBH}

\begin{document}
\makeatletter\renewcommand{\ps@plain}{%
	\def\@evenhead{\hfill\itshape\rightmark}%
	\def\@oddhead{\itshape\leftmark\hfill}%
	\renewcommand{\@evenfoot}{\hfill\small{--~\thepage~--}\hfill}%
	\renewcommand{\@oddfoot}{\hfill\small{--~\thepage~--}\hfill}%
}\makeatother\pagestyle{plain}


\title{Magnetic-field dependence of low-energy magnons, anisotropic heat conduction,\\ and spontaneous relaxation of magnetic domains in the cubic helimagnet ZnCr$_2$Se$_4$}

\author{D.~S. Inosov}\email[Corresponding author: ]{dmytro.inosov@tu-dresden.de}
\affiliation{Institut f{\"u}r Festk{\"o}rper- und Materialphysik, Technische Universit{\"a}t Dresden, 01069 Dresden, Germany}
\affiliation{W\"urzburg-Dresden Cluster of Excellence on Complexity and Topology in Quantum Matter\,---\,\textit{ct.qmat}, Technische Universit{\"a}t Dresden, 01069 Dresden, Germany}

\author{Y.~O. Onykiienko}
\affiliation{Institut f{\"u}r Festk{\"o}rper- und Materialphysik, Technische Universit{\"a}t Dresden, 01069 Dresden, Germany}

\author{Y.~V. Tymoshenko}
\affiliation{Institut f{\"u}r Festk{\"o}rper- und Materialphysik, Technische Universit{\"a}t Dresden, 01069 Dresden, Germany}

\author{A.~Akopyan}
\affiliation{Department of Physics, University of Miami, Coral Gables, Florida 33124, USA}

\author{D.~Shukla}
\affiliation{Department of Physics, University of Miami, Coral Gables, Florida 33124, USA}

\author{N.~Prasai}
\altaffiliation[Current address: ]{Department of Physics, St. Mary's College of Maryland, St. Mary's City, MD 20686, USA.}
\affiliation{Department of Physics, University of Miami, Coral Gables, Florida 33124, USA}

\author{M.~Doerr}
\affiliation{Institut f{\"u}r Festk{\"o}rper- und Materialphysik, Technische Universit{\"a}t Dresden, 01069 Dresden, Germany}
\affiliation{W\"urzburg-Dresden Cluster of Excellence on Complexity and Topology in Quantum Matter\,---\,\textit{ct.qmat}, Technische Universit{\"a}t Dresden, 01069 Dresden, Germany}

\author{D.~Gorbunov}
\affiliation{Hochfeld-Magnetlabor Dresden (HLD-EMFL), Helmholtz-Zentrum Dresden-Rossendorf, 01328 Dresden, Germany}

\author{S.~Zherlitsyn}
\affiliation{Hochfeld-Magnetlabor Dresden (HLD-EMFL), Helmholtz-Zentrum Dresden-Rossendorf, 01328 Dresden, Germany}

\author{D.\,J.~Voneshen}
\affiliation{ISIS Facility, STFC, Rutherford Appleton Laboratory, Didcot, Oxfordshire OX11-0QX, United Kingdom}
\affiliation{Department of Physics, Royal Holloway University of London, Egham, TW20-0EX, United Kingdom}

\author{M.~Boehm}
\affiliation{Institut Laue-Langevin, 71 avenue des Martyrs, CS 20156, 38042 Grenoble Cedex 9, France}

\author{V.~Tsurkan}
\affiliation{Experimental Physics V, Center for Electronic Correlations and Magnetism, Institute of Physics, University of Augsburg, 86135 Augsburg, Germany}
\affiliation{Institute of Applied Physics, Chisinau MD-2028, Republic of Moldova}

\author{V.~Felea}
\affiliation{Institute of Applied Physics, Chisinau MD-2028, Republic of Moldova}

\author{A.~Loidl}
\affiliation{Experimental Physics V, Center for Electronic Correlations and Magnetism, Institute of Physics, University of Augsburg, 86135 Augsburg, Germany}

\author{J.~L. Cohn}\email[Corresponding author: \vspace{-3pt}]{jcohn@miami.edu}
\affiliation{Department of Physics, University of Miami, Coral Gables, Florida 33124, USA}

\begin{abstract}\noindent\parfillskip=0pt\relax
Anisotropic low-temperature properties of the cubic spinel helimagnet ZnCr$_2$Se$_4$ in the single-domain spin-spiral state are investigated by a combination of neutron scattering, thermal conductivity, ultrasound velocity, and dilatometry measurements. In an applied magnetic field, neutron spectroscopy shows a complex and nonmonotonic evolution of the spin-wave spectrum across the quantum-critical point that separates the spin-spiral phase from the field-polarized ferromagnetic phase at high fields. A tiny spin gap of the pseudo-Goldstone magnon mode, observed at wave vectors that are structurally equivalent but orthogonal to the propagation vector of the spin helix, vanishes at this quantum critical point, restoring the cubic symmetry in the magnetic subsystem. The anisotropy imposed by the spin helix has only a minor influence on the lattice structure and sound velocity but has a much stronger effect on the heat conductivities measured parallel and perpendicular to the magnetic propagation vector. The thermal transport is anisotropic at $T\lesssim 2$~K, highly sensitive to an external magnetic field, and likely results directly from magnonic heat conduction. We also report long-time thermal relaxation phenomena, revealed by capacitive dilatometry, which are due to magnetic domain motion related to the destruction of the single-domain magnetic state, initially stabilized in the sample by the application and removal of magnetic field. Our results can be generalized to a broad class of helimagnetic materials in which a discrete lattice symmetry is spontaneously broken by the magnetic order.
\end{abstract}

\maketitle

\section{Introduction}

The family of chromium spinels with the general formula $A^{2+}$Cr$^{3+}_2X^{2-}_4$ ($A$~=~Mg, Zn, Cd, Hg\,---\,nonmagnetic metal ion; $X$~=~O, S, Se\,---\,chalcogen ion) is a perfect playground to investigate different types of noncollinear magnetic order resulting from frustrated magnetic interactions. The magnetic Cr$^{3+}$ ions with $S = \frac{3}{2}$ and $L = 0$ form an undistorted pyro\-chlore sublattice. Depending on the distance between Cr nearest
neighbors, the dominant nearest-neighbor exchange interaction $J_1$ may vary from antiferro- ($J_1 > 0$ in oxides) to ferromagnetic ($J_1 < 0$ in sulfides and selenides)~\cite{Yaresko08}.

On the one hand, a system of antiferromagnetically interacting classical Heisenberg spins on the pyrochlore lattice is geometrically frustrated and serves as an example of a spin-liquid system~\cite{MoessnerChalker98, BentonJaubert16}. However, \textit{ab-initio} calculations show that Cr-spinels also have non-negligible further-neighbor exchange interactions~\cite{Yaresko08}, which should generally reduce the frustration and stabilize long-range magnetic order~\cite{OkuboNguyen11}. On the other hand, if the nearest-neighbor interaction is ferromagnetic (FM), the frustration is relieved, and unlike in the antiferromagnetic (AFM) case, the introduction of further-neighbor exchanges can promote bond frustration in the system and lead to noncollinear ordered ground states. In a recent study, the classical phase diagram of the pyrochlore lattice with FM nearest-neighbor interactions was constructed taking into account exchanges up to the 4th nearest neighbor~\cite{TymoshenkoOnykiienko17}. It shows a rich variety of possible states depending on the relative strength of $J_2$ and $J_3$ exchange parameters, ranging from trivial collinear ferromagnetism to single- and multi-$q$ noncollinear spin structures.

Here we focus our attention on the ZnCr$_2$Se$_4$ compound\,---\,a typical member of the Cr-spinel family with FM nearest-neighbor interactions. It has a cubic crystal structure described by the space group $Fd\overline{3}m$ with the lattice constant $a=10.497$~\AA~\cite{Plumier66, HidakaTokiwa03}. Below \mbox{$T_\text{C} = 21$~K}, the Cr magnetic moments order in an incommensurate proper-screw structure with the propagation vector $\mathbf{q}_\text{h} = (0\,0\,q_\text{h})$, with $q_\text{h} \approx 0.44$~\cite{YokaichiyaKrimmel09, CameronTymoshenko16}. The high positive Curie-Weiss temperature $\Theta_\text{CW} = 90$--110~K~\cite{BaltzerLehmann65, HembergerNidda07} indicates the dominant role of relatively strong FM exchanges, although the presence of the competing interactions postpones the formation of long-range order. The distinctive feature of a frustrated system is the presence of short-range spin correlations way above the ordering temperature. In ZnCr$_2$Se$_4$, they were observed by $\mu$SR~\cite{ZajdelLi17} up to the Curie-Weiss temperature, which is around five times higher than the long-range ordering temperature. The negative thermal expansion that appears on the same temperature scale~\cite{HembergerNidda07} can also be attributed to the correlated spin fluctuations, implying the presence of significant spin-lattice coupling.

Specific-heat and thermal-expansion data show $\lambda$-like anomalies at $T_\text{C}$, which are much sharper than for conventional antiferromagnets, suggesting a weakly first-order transition~\cite{HembergerNidda07}, as it is usual for magnetic transitions coupled to a lattice instability~\cite{LarkinPikin69}. It is accompanied by a minute structural distortion to either tetragonal~\cite{KleinbergerKouchkovsky66, HidakaYoshimura03, HembergerNidda07, ChenYang14, ZajdelLi17} or orthorhombic~\cite{HidakaTokiwa03, YokaichiyaKrimmel09} structure with $(a-c)/a \approx 6\cdot10^{-4}$~\cite{ZajdelLi17}, most clearly evidenced by a strong splitting of certain phonon modes~\cite{RudolfKant07prb, RudolfKant07njp}. However, due to the underlying cubic symmetry, three equivalent magnetic domains are possible, with the propagation vectors pointing along three different cubic axes. In addition, two possible chiralities are equally likely within each domain. Therefore, if the sample is cooled down in the absence of external symmetry-breaking fields, magnetic domains with all possible ordering-vector orientations and both chiralities are equally populated, and the structural distortions are averaged out for a macroscopic sample.

The magnetic phase diagram of ZnCr$_2$Se$_4$ has been investigated with magnetization, ultrasound, and small-angle neutron scattering~\cite{FeleaYasin12, CameronTymoshenko16}. Upon application of a small magnetic field, the helical structure acquires a weak FM component along the field direction. After the field exceeds a certain critical value of approximately 1~T (at \mbox{$T=2$~K}, decreasing towards higher temperatures), a domain-selection transition takes place, favoring the helix with the smallest angle between its propagation vector and the field axis. It has been also demonstrated that the rotation of magnetic-field direction at low temperatures induces ferroelectric polarization, and a single-domain state with a unique chirality can be selected by the application of a sufficiently strong electric field in addition to the magnetic field~\cite{MurakawaOnose08, TokuraSeki10}.

With a further increase of magnetic field, the conical angle of the magnetic structure continues to decrease until it collapses to zero at the critical field value of $B_{\rm c} \approx 6.5$~T. According to the magnetization and ultrasound data, a small unsaturated spin component survives up to 10~T. Weak anomalies, observed at this field using both methods, lead to a suggestion about another high-field phase of so far unknown origin~\cite{FeleaYasin12}. While both domain-selection and the conical-to-collinear transition fields decrease upon warming, they are remarkably independent of the field direction. The ordering wave vector also turns out to be nearly insensitive to the magnetic field direction and strength, which means that the helical pitch length does not depend on the conical angle~\cite{CameronTymoshenko16}. In the latest study by Gu \textit{et al.}~\cite{GuZhao18}, the conical-to-collinear transition has been re-addressed from the point of view of field-driven quantum criticality, which is evidenced by the low-temperature $\propto T^2$ behavior of the specific heat and a sharp anomaly in thermal conductivity. In the updated phase diagram, the transition at $B_{\rm c}$ merges with the high-field transition to the field-polarized FM state in the limit $T \rightarrow 0$, leading to a well-defined quantum critical point (QCP) and a broad quantum-critical region at elevated temperatures~\cite{GuZhao18}.

In this work, we study the evolution of the low-energy magnon spectrum across the field-driven QCP and the anisotropic low-temperature heat conductivity in the single-domain helical phase. Using inelastic neutron scattering (INS), we uncover the mechanisms behind the formation of the quantum critical point associated with the collapse of the conical angle in the spin-spiral structure of a bond-frustrated helimagnet, which can be associated with field-driven softening and condensation of pseudo-Goldstone magnon modes and accidental restoration of cubic symmetry at the QCP. In line with our earlier prediction~\cite{TymoshenkoOnykiienko17}, we also demonstrate that the highly anisotropic and field-dependent low-energy magnon spectrum has a profound influence on the low-temperature thermal transport properties, resulting in a highly anisotropic heat conductance in an otherwise nearly cubic material. Further, we also investigate the stability of the single-domain helical state against thermal fluctuations. After such a state is stabilized in the sample by field-cooling or by application of a magnetic field above the domain-selection transition, as soon as the external magnetic field is removed, slow relaxation back to the multidomain state (on the scale of hours) is observed even at relatively low temperatures of about $0.5 T_\text{C}$.

\section{Sample preparation and experimental methods}

All experiments presented in this work have been carried out on high-quality single crystals of ZnCr$_2$Se$_4$ grown by chemical vapor transport. The crystal growth and characterization are described elsewhere~\cite{FeleaYasin12}. For the purpose of INS measurements, we used a mosaic sample consisting of 8 co-aligned single crystals with a total mass of approximately 1~g (the same as in Ref.~\cite{TymoshenkoOnykiienko17}). The crystals were oriented and co-aligned with an accuracy of about 1$^\circ$ using an x-ray Laue backscattering camera.

The \emph{INS measurements} were carried out using both time-of-flight (TOF) and triple-axis (TAS) spectrometers, equipped with vertical-field cryomagnets. The sample was aligned with its [001] axis vertical, so that the application of magnetic field promotes magnetic domains with the propagation vector $(0\,0\,q_\text{h})$ orthogonal to the scattering plane, thereby enabling INS measurements of the pseudo-Goldstone magnon modes at the orthogonal wave vectors $(q_\text{h}\,0\,0)$ and $(0\,q_\text{h}\,0)$ in the single-domain state. The TOF experiment~\cite{DataLET} was carried out at LET (ISIS)~\cite{BewleyTaylor11} with the incident neutron energy $E_\text{i} = 2$~meV, which yields the energy resolution of 0.04~meV at zero energy transfer. We collected the TOF data at 0, 3, 6, and 8~T. The advantage of the TOF method is that it allows us to cover a large volume of the energy-momentum space in a single measurement. To fill the gaps in the field dependence, we complemented the measurements at the TAS instrument ThALES (ILL)~\cite{BoehmHiess08, BoehmSteffens15}, focusing our attention only at particular points in reciprocal space~\cite{DataThales}. The instrument was operated with a fixed final neutron wave vector, $k_\text{f} = 1.1$~\AA\ (2.5~meV). Pyrolytic-graphite monochromator and analyzer were used in the double-focusing mode to increase the effective neutron flux, and a cold beryllium filter was installed between the sample and analyzer to reduce the contamination with higher-order neutrons from the monochromator. For selected scans, we also installed a collimator to decrease the background and improve the~resolution.

For \emph{ultrasound measurements}, we polished two opposite $\{100\}$ surfaces of a single crystal. Piezoelectric transducers were glued to the polished surfaces to induce and detect ultrasound vibrations with the generation frequency of 100~MHz and longitudinal polarization. In this configuration, both the wave vector of the ultrasound, $\mathbf{k}_\text{s}$, and its polarization were parallel to the [100] cubic axis of the crystal. The same sample was measured with two orientations of the magnetic field: $\mathbf{B} \parallel [100] \parallel \mathbf{k}_\text{s}$ and $\mathbf{B} \parallel [001] \perp \mathbf{k}_\text{s}$, which are equivalent in the cubic setting above $T_{\rm C}$.

The same polished single crystal was then used to measure \textit{dilatometric properties} of the sample. We used a tilted-plate capacitive dilatometer with a sensitivity to relative length changes of about $10^{-7}$~\cite{RotterMueller98} to assess relative changes in the lattice constants as a function of magnetic field and temperature. The thermal expansion between 1.5~K and 300~K and magnetostriction in steady fields were measured using a superconducting 15~T magnet (Oxford Instruments) at the TU Dresden. To determine the relaxation process from a single-domain to multidomain state, the sample was first cooled down in a field of 7~T, then the field was ramped down with the maximal rate of 1~T/min, and as soon as it reached zero, the dilatometer signal was measured over a long time.

Specimens for \textit{thermal-conductivity measurements} were cut from as-grown single crystals, oriented by x-ray diffraction, and polished into thin parallelepipeds with edges along $\langle 100\rangle$ directions. A two-thermometer\,--\,one-heater method was employed to measure the thermal conductivity using a matched pair of RuO sensors thermally linked to the specimens with $\varnothing0.005''$ gold wires bonded with silver epoxy. The sensors were calibrated in each experiment against a calibrated reference sensor mounted in the Cu heat sink. Experiments with the applied field perpendicular and parallel to the heat flow were performed in successive low-temperature runs using the same thermometry attachments. The accuracy of the measurements is limited by uncertainty in the geometric factor to approximately 10\%.

\vspace{-3pt}\section{Anisotropy in the lattice parameters and sound velocities}\vspace{-2pt}
\label{Sec:Anisotropy}

Our subsequent discussion substantially relies on the understanding of spin-lattice coupling and the origins of spin anisotropy in the magnetically ordered state. On the one hand, in the high-temperature cubic phase, the Cr~$t_{2g}$ shell is half-filled, quenching the orbital moment to zero. Thus, no on-site spin-orbit interaction is expected. On the other hand, partial transfer of the magnetic moment to the neighboring Se sites was suggested based on recent nuclear magnetic resonance (NMR) measurements~\cite{ParkKwon19}. This mechanism can potentially induce finite orbital component and, therefore, facilitate the spin-lattice interaction. The theory confirms this possibility, but the corresponding corrections are expected to be insignificant~\cite{Yaresko08}.

The response of the lattice to the onset of helimagnetic order is manifested in the tetragonal or orthorhombic lattice distortion below $T_{\rm C}$, which is mainly due to the magnetostrictive displacements of Se$^{2-}$ ions resulting from the magnetic interaction of neighboring Cr$^{3+}$ spins~\cite{HidakaTokiwa03}. According to various estimates, this distortion does not exceed 0.06--0.08\% at the lowest measured temperature~\cite{KleinbergerKouchkovsky66, HidakaYoshimura03, HembergerNidda07, ZajdelLi17}. In view of the small spin-orbit coupling, any anisotropic terms in the spin Hamiltonian that may result from this minor deviation from the cubic symmetry are usually considered negligible, leading to the Heisenberg-type behavior of magnetic moments with no detectable spin-space anisotropy.

Structural distortions of the order of 10$^{-4}$ are at the limit of sensitivity for conventional diffraction methods. Therefore, in our present work we used capacitive dilatometry with a sensitivity of $\sim$10$^{-7}$ to measure relative changes in the lattice constants more accurately as a function of magnetic field and temperature. The sample was mounted in the dilatometer to measure relative changes of its length, $\Delta l/l_0$, along one of its $\langle 001 \rangle$ cubic axes. Depending on the direction of magnetic field (transverse or longitudinal with respect to the chosen axis), it corresponds to either $a$ or $c$ axis, respectively, in the tetragonally distorted structure with $a = b \neq c$. Therefore, we can measure relative changes in both $a$ and $c$ lattice constants on the same single crystal without remounting it inside the dilatometer. All measurements are presented relative to the $l_0$ value, corresponding to the high-temperature cubic structure at $B=0$ and $T=30$~K.

\begin{figure}[t]
\includegraphics[width=\linewidth]{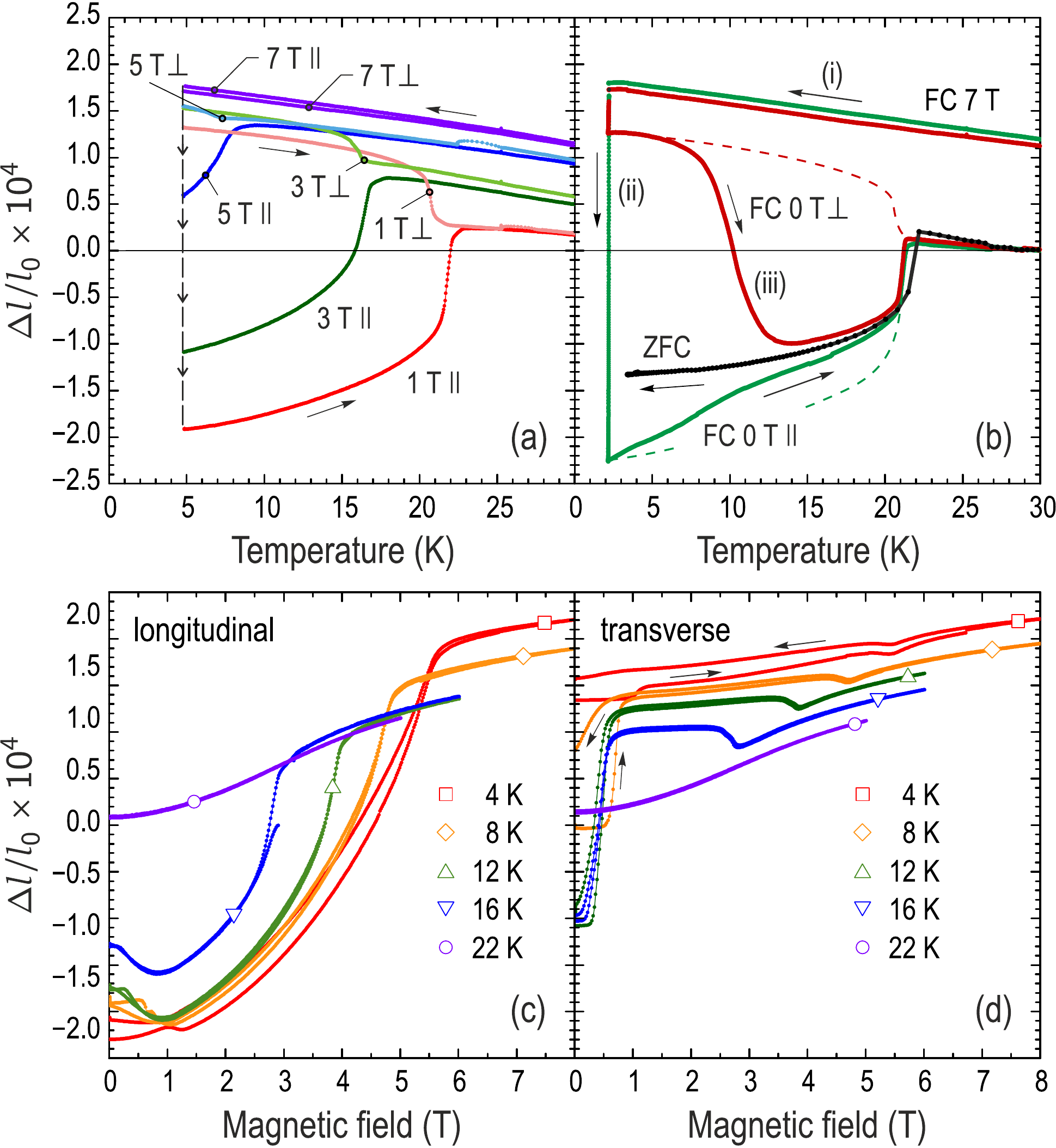}\vspace{-3pt}
\caption{(a)~Thermal-expansion measurements in longitudinal and transverse magnetic fields above the domain-selection transition. All curves are normalized to the same $l_0$ value at $T=30$~K and $B=0$. (b)~Zero-field thermal expansion, measured during zero-field cooling (ZFC) and after field-cooling in a longitudinal (``FC 0~T $\parallel$'') or transverse (``FC 0~T $\perp$'') field of 7~T. The measurement steps (i), (ii), (iii) are explained in the text. (c,\,d)~Magnetostriction measurements at several constant temperatures for longitudinal and transverse field directions, respectively.\vspace{-3pt}}
\label{Fig:Dilatometry}
\end{figure}

Figure \ref{Fig:Dilatometry}\,(a) shows the temperature dependence of $\Delta l/l_0$, measured in longitudinal and transverse magnetic fields ($B=1$, 3, 5, and 7~T) above the domain-selection transition. In all measurements with $B \geq 1$~T, the sample always remains in the single-domain state. The longitudinal and transverse thermal-expansion curves essentially coincide at high temperatures (\mbox{$T > T_\text{C}$}) or in high magnetic fields above the saturation ($B > B_\text{c}$), suggesting that both para\-magneto\-striction (forced magnetostriction above the ordering temperature) and thermal expansion in the field-polarized FM state are isotropic. This behavior is typical for systems with spin-only magnetic moments~\cite{TakahashiShimizu77, TremoletdeLacheisserie93, CullityGraham09}, in contrast to strongly anisotropic magnetostriction in some rare-earth local-moment systems that feature an aspherical 4$f$ electron shell with a nonzero orbital angular momentum~\cite{CroftZoric78, ZieglowskiWohlleben87, FawcettPluzhnikov91, SokolovNakamura99}. Spontaneous magnetostriction in the helimagnetic state, on the contrary, is very anisotropic, resulting in the splitting of the thermal-expansion curves below $T_\text{C}$. The lattice contracts in the direction parallel to the spin spiral and expands in the orthogonal directions, which results in a tetragonal distortion with $(a-c)/a \approx 3.6\cdot10^{-4}$ in the zero-field and zero-temperature limit, that is even smaller than in earlier estimates from x-ray diffraction~\cite{KleinbergerKouchkovsky66, HidakaYoshimura03, ChenYang14, ZajdelLi17}.

The zero-field thermal expansion, measured after zero-field cooling (ZFC) and after field-cooling (FC) in longitudinal and transverse magnetic fields of 7~T, are shown separately in Fig.~\ref{Fig:Dilatometry}\,(b) because they show qualitatively different behavior. The ZFC reference curve (black), which represents the thermal expansion of ZnCr$_2$Se$_4$ in the multi-domain state, is in perfect agreement with earlier measurements by Gu \textit{et al.}~\cite{GuYang16}. (i)~Because the field cooling takes place above the saturation field, where the magnetostriction is isotropic, the field does not break the cubic symmetry of the lattice, and the corresponding transverse and longitudinal thermal-expansion curves (labeled ``FC 7~T'') coincide. (ii)~Then, at the base temperature of 2~K, the magnetic field is ramped down to zero with a rate of 0.4~T/min, wherein the sample undergoes a transition to the helimagnetic tetragonal state with $\mathbf{c}\parallel\mathbf{q}_\text{h}\parallel\mathbf{B}$. In the longitudinal geometry ($c$ axis), where the forced and spontaneous contributions to the magnetostriction have the same sign, this results in a significant contraction of the sample with a change in $\Delta l/l_0$ of approximately $4\times10^{-4}$, which appears in Fig.~\ref{Fig:Dilatometry}\,(b) as a vertical line. In the transverse geometry ($a$ axis), the forced and spontaneous contributions nearly cancel each other, so that the resulting change in $\Delta l/l_0$ is only $0.5\times10^{-4}$. (iii)~After stabilizing the single-domain state of the sample by field cooling, thermal expansion is measured in zero field upon increasing the temperature. One could expect that the single-domain state survives up to $T_\text{C}$, which would result in the order-parameter-like change in both lattice parameters shown with dashed lines, similar to that observed in finite fields [Fig.~\ref{Fig:Dilatometry}\,(a)]. However, measured thermal-expansion curves (labeled ``FC 0~T $\parallel$'' and ``FC 0~T $\perp$'') deviate from this behavior at temperatures above 5~K, as they both converge to the ZFC reference measurement. This suggests that the sample relaxes back to the multidomain state at elevated temperatures on the time scale of our measurement, which was carried out with a temperature sweep rate of 0.1 K/min. The result is most pronounced in the transverse geometry, resulting in a nonmonotonic sign-changing behavior of the magnetostrictive strain.

The slow magnetic relaxation to the multi-domain state is also revealed in the measurements of magnetostriction at several constant temperatures, which are presented in Figs.~\ref{Fig:Dilatometry}\,(c,\,d) for the longitudinal and transverse field geometries, respectively. All measurements start in the single-domain state achieved by cooling the sample in a 3~T field. The field is then swept to the maximal field and back to zero, which in the absence of relaxation processes should result in identical magnetostriction curves. One can see, however, that already at 4~K the curves slightly deviate from each other, and a weak anomaly is observed at the domain-selection transition around 1~T, which is not expected if the sample remained in the single-domain state after the field was switched off. This anomaly becomes much more pronounced at 8~K, where the domain relaxation is faster, although the curves measured upon increasing and decreasing the field still show a hysteresis, suggesting that the sample does not fully relax back to the multidomain state on the time scale of the measurement. Most remarkably, a downturn (upturn) in the transverse (longitudinal) magnetostriction measured in decreasing field starts already at 0.8~T, which implies that the domain relaxation takes place not only in zero field but also in finite fields as long as they remain below the domain-selection transition. At higher temperatures of 12 or 16~K, approaching $T_\text{C}$, the relaxation becomes so fast that the magnetostriction curves measured in upward and downward field sweeps nearly coincide. Finally, the measurements at \mbox{$T=22$~K~>\,$T_\text{C}$} reveal nearly isotropic paramagnetostriction, which remains quadratic in field up to approximately 2~T with $\epsilon(B)/B^2\approx7\cdot10^{-6}$~T$^{-2}$, where $\epsilon(B)=\left[l(B,T)-l(0,T)\right]/l(0,T)$ is the parastrictive strain.

\begin{figure}[b]
\centerline{\includegraphics[width=0.97\linewidth]{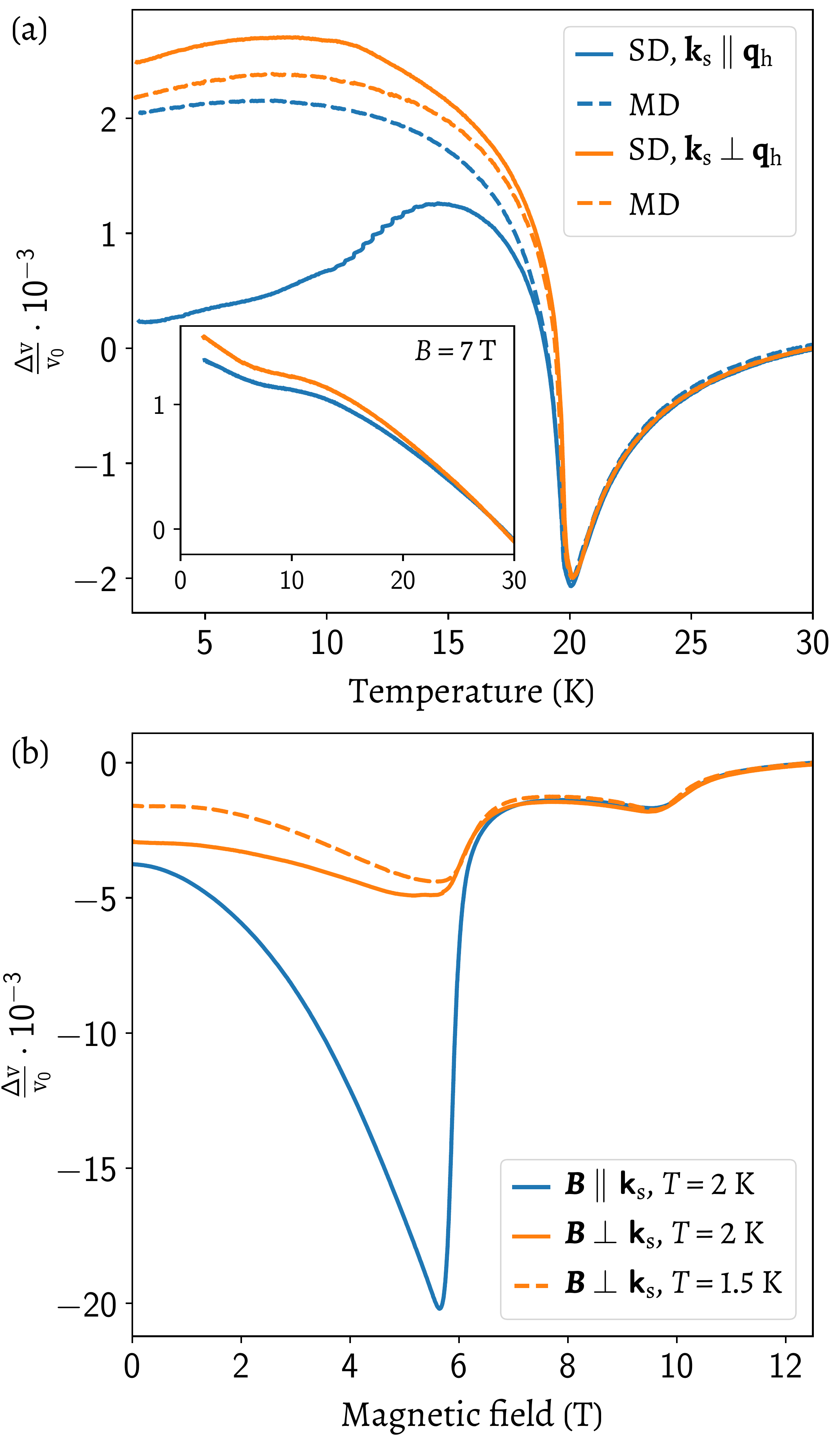}}\vspace{-2pt}
\caption{(a)~The change in the sound velocity of the longitudinal acoustic mode, $\Delta v=v-v_0$, normalized to the value $v_0$ at 30~K, in the single-domain (SD, solid lines) and multidomain (MD, dashed lines) states. The top and bottom curves correspond to the ultrasound wave vector $\mathbf{k}_\text{s}$ perpendicular and parallel to the propagation vector of the spin helix, $\mathbf{q}_\text{h}$, respectively. The inset shows a similar temperature dependence, measured in the field-polarized FM phase at $B=7$~T for $\mathbf{k}_\text{s}\perp\mathbf{B}$ (top curve) and $\mathbf{k}_\text{s}\parallel\mathbf{B}$ (bottom curve), for comparison. (b)~Magnetic-field dependence of the relative change in the sound velocity, $\Delta v/v_0$, normalized to the respective $v_0$ values at 12~T. The measurements are done at a constant temperature with the ultrasound wave vector parallel and perpendicular to the magnetic-field direction. Note the tenfold difference in the ranges of the vertical scales between (a) and (b).\vspace{-2pt}}
\label{Fig:Ultrasound}
\end{figure}

Before moving on to a more detailed consideration of the slow domain relaxation and discussing the possible mechanisms of such behavior, we would like to quantify the lattice anisotropy in the helimagnetic state of ZnCr$_2$Se$_4$ using a complementary approach that consists in comparing sound velocities in the directions parallel and orthogonal to the magnetic propagation vector. The ultrasound data in Fig.~\ref{Fig:Ultrasound}\,(a) show relative changes in the sound velocity, $\Delta v/v_0$, for the longitudinal wave propagating along the $[001]\parallel\mathbf{q}_\text{h}$ and $[100]\perp\mathbf{q}_\text{h}$ directions, normalized to the respective values in the paramagnetic state at $T=30$~K. It is clear that the crystal is slightly stiffer in the direction orthogonal to the ordering vector of the spin helix, but the effect is relatively small, on the order of 0.2\%. This suggests that the difference in phonon energies and phononic densities of states along the two directions must also be similarly small. For comparison, the inset to Fig.~\ref{Fig:Ultrasound}\,(a) shows a similar temperature dependence, measured in the field-polarized FM phase at $B=7$~T in the directions parallel and orthogonal to the field-induced magnetization. The two curves have been normalized to the respective $v_0$ values at $T=30$~K, where the sample remains cubic according to the paramagnetostriction measurements described above. The observed change in $\Delta v/v_0$ between 30~K and base temperature differs by only $2\cdot10^{-4}$ for the two directions of the ultrasound wave vector, which is within the reproducibility margin of our measurements.

The magnetic-field dependence of the relative change in the sound velocity, $\Delta v/v_0$, measured for the same two directions at $T=2$~K and normalized to the high-field value at 12~T, is plotted in Fig.~\ref{Fig:Ultrasound}\,(b). One can see a very pronounced softening of the longitudinal acoustic mode immediately below the critical field, which is much stronger for the acoustic waves propagating along the magnetic-field direction than perpendicular to it. The maximal value of $|\Delta v/v_0|$ reaches 2\%, which is ten times larger than the corresponding changes at base temperature and almost two orders of magnitude larger than the spontaneous magnetostriction, suggesting that magnetic interactions exert a much stronger relative effect on the dynamic than static properties of the lattice. The second anomaly that appears in both field-dependence curves around 10~T represents the previously reported high-field transition that was previously associated either with a spin-nematic phase~\cite{FeleaYasin12} or with a crossover from the quantum-critical region to the fully field-polarized FM phase~\cite{GuZhao18}. The new phase diagram suggested by Gu \textit{et al.}~\cite{GuZhao18} implies a very strong temperature dependence of this high-field anomaly, so that it merges with $B_\text{c}$ in the zero-temperature limit. In contrast, we can see essentially no difference between the 1.5 and 2~K curves in Fig.~\ref{Fig:Ultrasound}\,(b), indicating that the crossover to the field-polarized FM phase stays at approximately 10~T at both temperatures, which is hard to reconcile with the newly proposed phase~diagram.

\section{Slow relaxation of helimagnetic domains}
\label{Sec:Relax}

\begin{figure}[t]
\includegraphics[width=\linewidth]{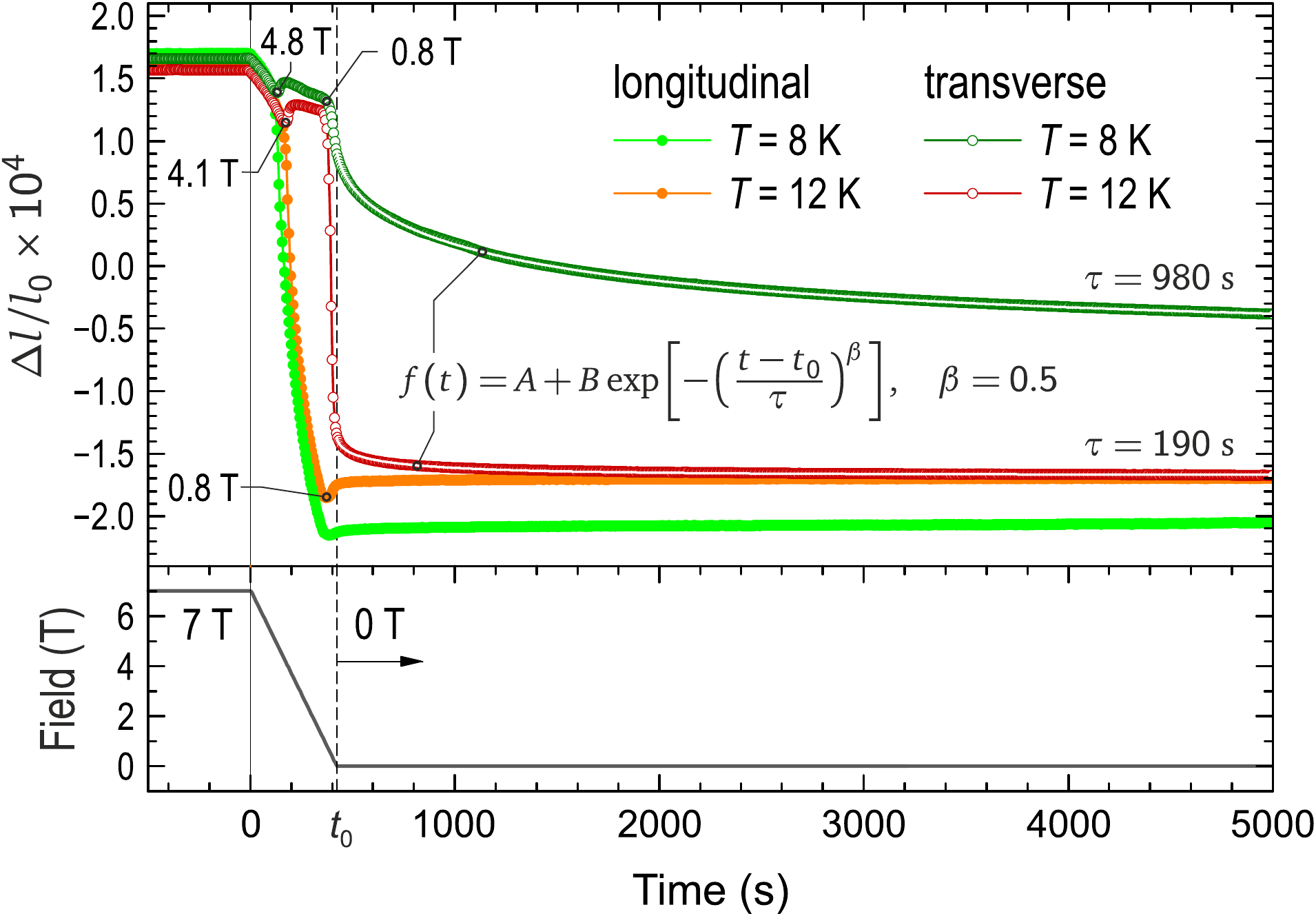}\vspace{-3pt}
\caption{Time dependence of the relative change in the sample length $\Delta l/l_0$, measured during and after a rapid linear sweep of magnetic field from 7~T to zero in the transverse and longitudinal directions. The time dependence of magnetic field is plotted at the bottom. At $t<0$, the sample temperature is stabilized at $T=8$ or 12~K, and a downward field sweep with 1~T/min starts. After the time $t_0=420$~s, when the field reaches zero, slow relaxation towards the multidomain state can be described by a stretched exponential, Eq.~(\ref{Eq:StretchedExp}), with $\beta=0.5$ and the average decay constants $\tau_\text{8\,K}=980$~s and $\tau_\text{\!12\,K}=190$~s (thin white lines).\vspace{-3pt}}
\label{Fig:Relaxation}
\end{figure}

Having observed slow magnetic domain relaxation in magnetostriction measurements, we proceed with estimating the functional dependence and characteristic times of such relaxation by following the time-dependent changes in the sample length at a constant temperature. After applying a 7~T field to the sample in either longitudinal or transverse direction and stabilizing the temperature at $T=8$ or 12~K, we swept the field down to zero with the maximal rate of 1~T/min and followed the time-dependent change in the relative sample length, $\Delta l/l_0$, both during and after the field sweep. The results are plotted in Fig.~\ref{Fig:Relaxation}, where the bottom panel shows the time dependence of the external field. During the downward field sweep, anomalies in $\Delta l/l_0$ are observed at the critical field (4.8 or 4.1~T, respectively) and at the domain-selection field (0.8~T). As soon as the magnetic field is reduced below the domain-selection field, slow relaxation towards the multidomain state is observed both in the transverse and longitudinal directions. The relaxation curves cannot be described by a simple exponential function, which implies contributions from processes with a broad spectrum of different relaxation times. Defining the time when magnetic field reaches zero as $t_0$, we can describe the relaxation at times $t>t_0$ using the stretched exponential function,\vspace{-6pt}
\begin{equation}\label{Eq:StretchedExp}
f(t)=A + B\,\exp\left[-\left(\frac{\displaystyle t-t_0}{^{\phantom{0}}\displaystyle\tau~~}\right)^\beta\right],\vspace{-3pt}
\end{equation}
with $\beta = 0.5$, where $\tau$ is the temperature-dependent decay constant. This model is equivalent to a continuous weighted sum of simple exponential decays with a distribution of relaxation times described by the probability density\vspace{-3pt}
\begin{equation}\label{Eq:StretchedExpDistr}
\rho(\theta)=\frac{1}{2\sqrt{\pi}}\theta^{-1/2}\exp(-\theta/4).\vspace{-3pt}
\end{equation}

Relaxation with a stretched-exponential behavior is typical for systems with randomly distributed magnetic domains and is commonly associated with magnetic domain-wall motion. Here it results in the destruction of the single-domain state on the time scale of several minutes or hours (depending on the temperature) with a tendency to restore the cubic symmetry of the sample on the macroscopic scale. At a lower temperature of 8~K, the average relaxation rate is approximately 5 times slower than at 12~K. Apart from that, the two datasets also show a qualitative difference. At the higher temperature of 12~K, both transverse and longitudinal curves relax to the same asymptotic value of $\Delta l/l_0$ at \mbox{$t\rightarrow\infty$}, suggesting that an isotropic multidomain state is fully restored as a result of such relaxation. In contrast, the 8~K datasets relax to different asymptotic values in the transverse and longitudinal directions, which implies that some partial anisotropy in the distribution of magnetic domains survives arbitrarily long after the external field is turned off. A crossover between these two distinct behaviors must therefore take place at some intermediate temperature between 8 and 12~K.

Because every domain wall in a helimagnet costs additional energy, relaxation through nucleation of additional misaligned domains is only possible if some excess energy is already present in the initial state of the system. Indeed, by applying magnetic field to the sample, we only select magnetic domains with a single direction of the magnetic ordering vector, but such domains can still possess different chiralities. One therefore expects that domain walls separating right- and left-handed helical domains are still present in the initial state, carrying excess energy that can be redistributed via nucleation of new magnetic domains with a misaligned orientation of the magnetic ordering vector as a result of relaxation processes facilitated by thermal fluctuation. In other words, a domain wall separating helimagnetic domains with opposite chirality can decay into a larger number of domain walls between domains with orthogonal ordering vectors and either equal or opposite chiralities. It is natural to expect that such processes are governed by the corresponding domain-wall energies that must be different for all three types of domain walls and could be estimated from micromagnetic simulations.

\section{Inelastic neutron scattering}
\label{Sec:INS}

\begin{figure*}[t]
\includegraphics[width=\linewidth]{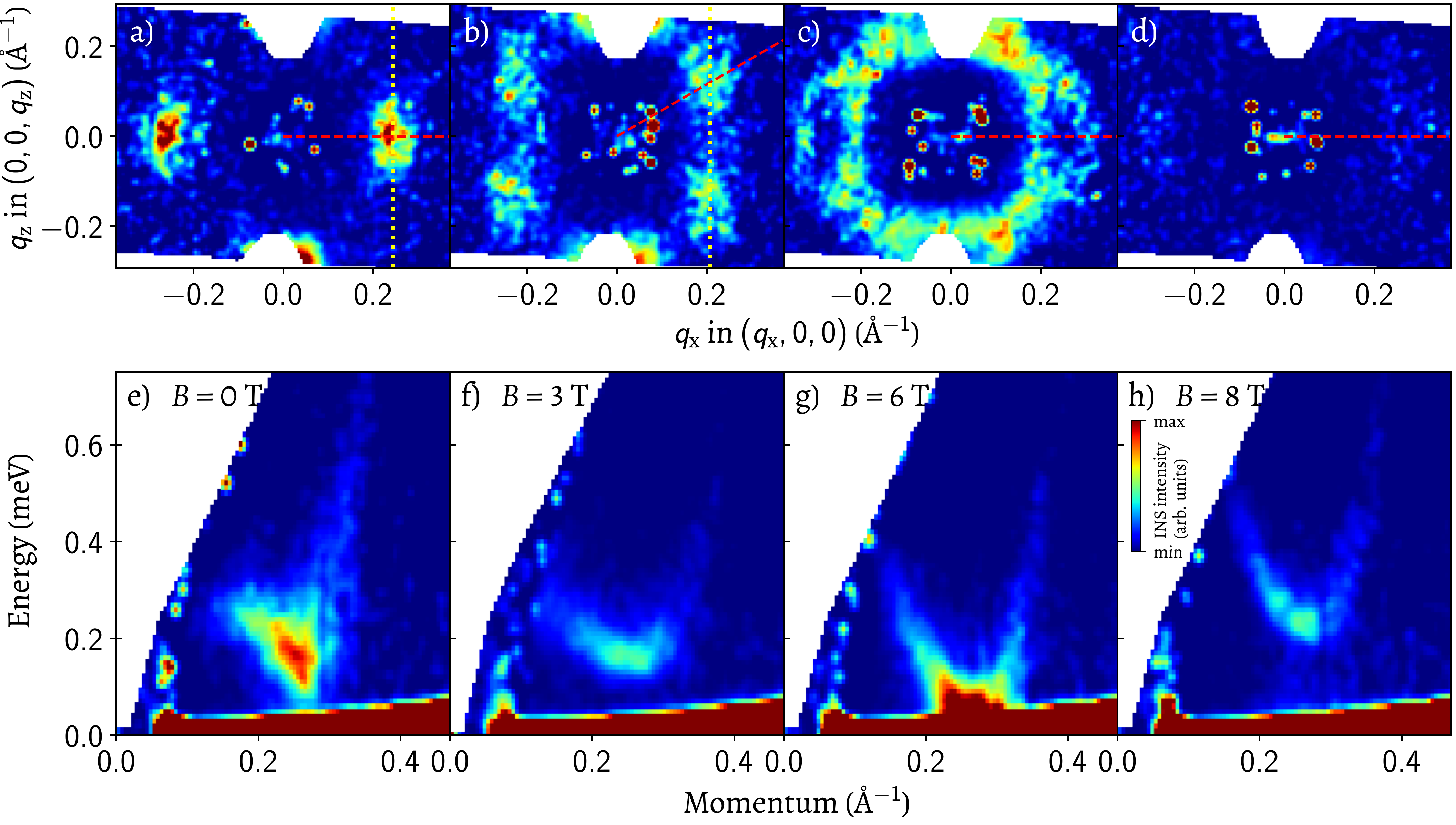}\vspace{-3pt}
\caption{The neutron TOF spectrum of ZnCr$_2$Se$_4$, measured in magnetic fields of 0, 3, 6, and 8~T, applied along the [001] direction. The top row of panels (a--d) shows the corresponding constant-energy cuts in the $(H0L)$ plane, integrated within $\pm 0.03$~meV around 0.17~meV energy transfer. The bottom row (e--h) shows energy-momentum cuts along the radial directions connecting the $\Gamma$ point with the local minimum of the dispersion, as shown with dashed lines in (a--d). Integration ranges in momentum directions orthogonal to the plane of the figure are approximately $\pm 0.05$~\AA$^{-1}$. The vertical dotted lines in (a) and (b) indicate the direction of energy-momentum cuts in Fig.~\ref{fig:TOF_trans}.}
\label{fig:TOF}
\end{figure*}

The main goal of our INS measurements is to investigate the magnetic-field dependence of the previously discovered pseudo-Goldstone magnon modes in ZnCr$_2$Se$_4$~\cite{TymoshenkoOnykiienko17} and reveal their relationship to the field-driven QCP suggested by Gu \textit{et al.}~\cite{GuZhao18}. In the paramagnetic phase above $T_{\rm C}$, the three pairs of wave vectors \mbox{$(\pm q_\text{h}\,0~0)$}, \mbox{$(0\,\pm\!q_\text{h}\,0)$}, and \mbox{$(0~0\,\pm\!q_\text{h})$} are equivalent because of the cubic symmetry. Helimagnetic order breaks this discrete symmetry by spontaneously selecting one of the cubic directions as the propagation direction of the spin spiral. For the sake of unambiguity, let us assume that this direction is \mbox{$(0~0\,\pm\!q_\text{h})$}. Consequently, all spins forming the helix are constrained to the $ab$ plane. Under the assumption of Heisenberg symmetry, which holds to a very good approximation for Cr$^{3+}$ spins, the system remains invariant under continuous $U(1)$ spin rotations within the $ab$ plane. As a consequence, Goldstone modes emanating from the \mbox{$(0~0\,\pm\!q_\text{h})$} ordering vectors remain gapless down to the lowest temperatures (the upper limit for the spin gap at these points, set by previous INS experiments, is 0.05~meV~\cite{ZajdelLi17}).

Under the assumption that the lattice symmetry remains cubic, so that no spin-space anisotropy is imposed by the lattice distortions, linear spin-wave theory (LSWT) predicts gapless magnon modes also at the two orthogonal pairs of wave vectors, \mbox{$(\pm q_\text{h}\,0~0)$} and \mbox{$(0\,\pm\!q_\text{h}\,0)$}, which can be seen as an accidental degeneracy, because in contrast to the true Goldstone modes, the zero spin gap at these points is not enforced by the symmetry of the Hamiltonian. As a consequence, the degeneracy can be weakly lifted by fluctuations, opening up a small magnon gap at the corresponding wave vectors~\cite{RauMcClarty18} and resulting in pseudo-Goldstone modes that have been experimentally observed in ZnCr$_2$Se$_4$ by neutron spectroscopy~\cite{TymoshenkoOnykiienko17}. In the case of continuous degeneracies, this phenomenon is known as ``order by disorder''~\cite{VillainBidaux80, Henley89}, whereas in ZnCr$_2$Se$_4$ the degeneracy is discrete, resulting in one pair of Goldstone and two pairs of pseudo-Goldstone modes, well separated in the Brillouin zone.

According to our spin-wave calculations in zero magnetic field, quantum-fluctuation corrections to the spin-wave spectrum are qualitatively similar to those resulting from LSWT in the presence of weak easy-plane anisotropy (in the plane orthogonal to the spiral direction). Both effects break the cubic symmetry of the magnon dispersions by opening a gap at the pseudo-Goldstone wave vectors while keeping the true Goldstone modes gapless. In the first case, this symmetry breaking is spontaneous, whereas in the second case a preferred axis is explicitly introduced in the magnetic Hamiltonian to reflect the tetragonal distortion resulting from the spin-lattice coupling. On the one hand, experimentally observed pseudo-Goldstone magnon gap $\Delta_\text{PG} \approx 0.17$~meV shows reasonable agreement with the theoretical estimate that takes into account quantum fluctuations (three- and four-magnon processes) even in the cubic setting~\cite{TymoshenkoOnykiienko17}. This offers solid evidence that the anisotropy in the spin-wave spectrum results mainly from quantum-fluctuation corrections beyond LSWT. On the other hand, small easy-plane anisotropy cannot be fully excluded in the presence of a tetragonal lattice distortion, even if our results outlined in section~\ref{Sec:Anisotropy} indicate that the corresponding distortions are minor, and due to the smallness of spin-orbit coupling, the resulting spin anisotropy should be negligible. At the qualitative level, the contribution of magnetostrictive distortions to the easy-plane anisotropy and, consequently, to the pseudo-Goldstone gap energy remains unknown. Estimating it from first principles would require more elaborate relativistic calculations, taking into account both lattice and spin degrees of freedom, which are currently beyond reach.

To shed light on this problem, we have measured the low-energy magnon spectrum of ZnCr$_2$Se$_4$ in magnetic fields applied along the [001] cubic axis. Figure~\ref{fig:TOF} shows representative cuts through the TOF data at 0, 3, 6, and 8~T. The top row of panels [Fig.~\ref{fig:TOF}\,(a--d)] presents constant-energy cuts in the $(H0L)$ plane at 0.17~meV. This energy corresponds to the pseudo-Goldstone magnon gap in zero field~\cite{TymoshenkoOnykiienko17}. The red dashed lines connect the center of the Brillouin zone with the local minimum in the dispersion. In the panels below [Fig.~\ref{fig:TOF}\,(e--h)], corresponding momentum-energy cuts along these lines are shown.

\begin{figure}[b]
\includegraphics[width=\linewidth]{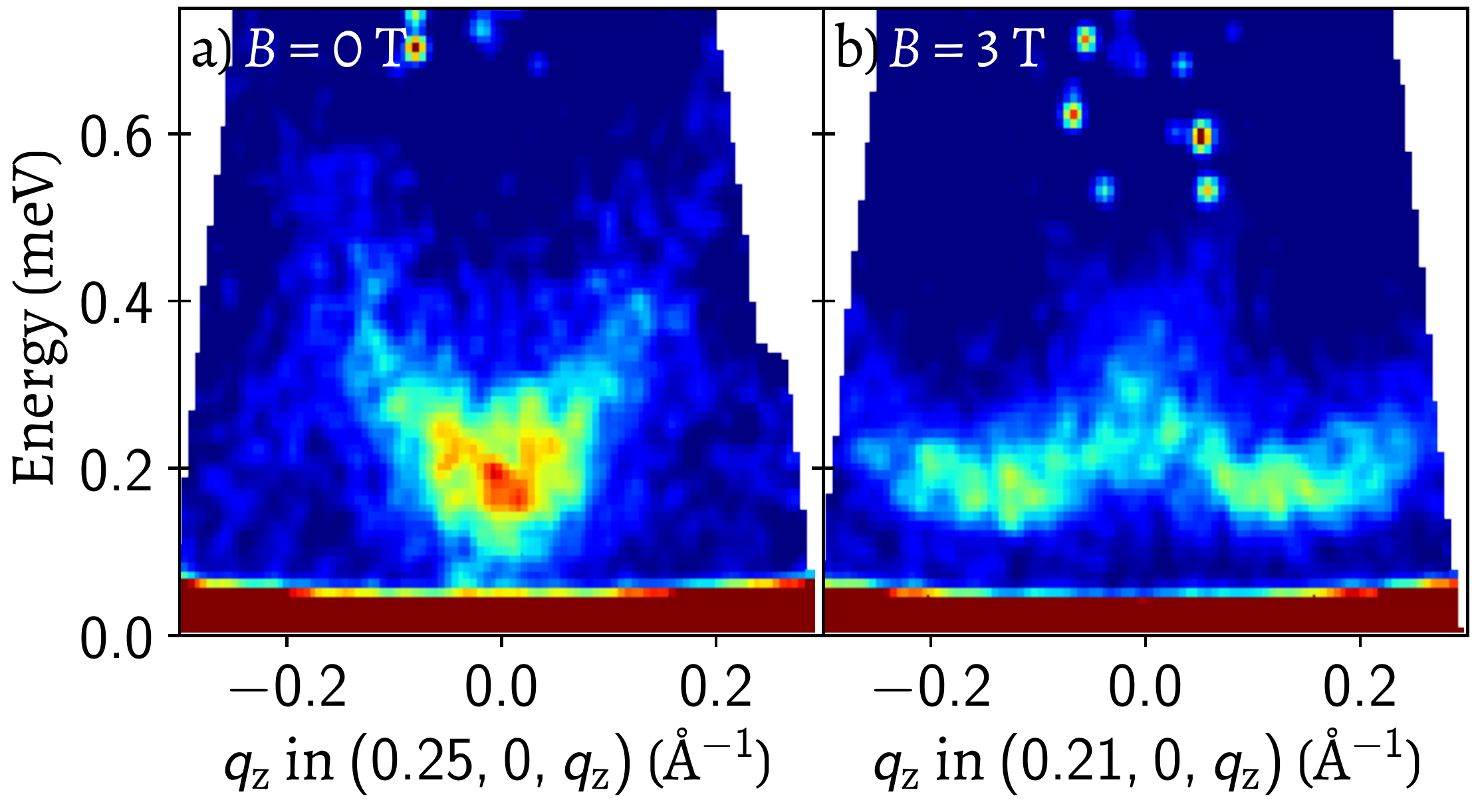}\vspace{-3pt}
\caption{Energy-momentum cuts along the $q_z$ direction in the $B=0$ and 3~T datasets, passing through the minima in the spin-wave dispersion, as indicated in Figs.~\ref{fig:TOF}\,(a,\,b), respectively, with vertical dotted lines. The integration range in $q_{\smash{x}}$ and $q_{\smash{y}}$ momentum directions is approximately $\pm 0.05$~\AA$^{-1}$.\vspace{-3pt}}
\label{fig:TOF_trans}
\end{figure}

The spectrum in Figs.~\ref{fig:TOF}\,(a,\,e) was measured in zero field. In agreement with our previous report~\cite{TymoshenkoOnykiienko17}, we see clear intensity maxima at ($\pm\!q_\text{h}$\,0~0), near the bottom of the pseudo-Goldstone modes. Their positions are equivalent (in the cubic setting) to the ordering vector, which is oriented along the $\mathbf{c}$ axis and cannot be reached in this experimental configuration. The dispersion has a sharp minimum, which is asymmetric even in the direct neighborhood of the ordering vector. On the left side from the minimum, closer to the $\Gamma$ point, the dispersion forms an arc with a maximum around 0.3~meV. On the right-hand side, the magnon branch rises linearly in energy and has a much steeper dispersion, forming a beak-like shape. The minimum of the dispersion corresponds to the pseudo-Goldstone gap energy. In the constant-energy plane in Fig.~\ref{fig:TOF}\,(a), the magnetic signal appears in the form of sharp spots at ($\pm\!q_\text{h}$\,0~0), with no additional intensity maxima in the field of view.

The application of a 3~T magnetic field, approximately half way to the QCP, qualitatively changes the shape of the dispersion. In the constant-energy plane in Fig.~\ref{fig:TOF}\,(b), the intensity now splits into four spots, which indicates that the dispersion minima are shifted up and down with respect to the $(H\,K\,0)$ plane and no longer coincide with the wave vectors equivalent to the magnetic ordering vector in the cubic setting. The momentum-energy cut through one of these minima in the dispersion, shown in Fig.~\ref{fig:TOF}\,(f), reveals a markedly different spectrum to the beak-shaped one in zero field. Even though the spin gap remains nearly unchanged, the dispersion no longer has a cusp and lacks the linear part on the right-hand side, approaching a parabolic shape. The comparison of 0 and 3~T cuts in Fig.~\ref{fig:TOF_trans}, taken through the corresponding minima in the dispersion along the transverse momentum direction [vertical dotted lines in Fig.~\ref{fig:TOF}\,(a,\,b)], indicates that the pseudo-Goldstone mode splits along the magnetic-field direction into two identical soft modes with a parabolic dispersion. The appearance of a crossing point of the two dispersion branches at the center of Fig.~\ref{fig:TOF_trans}\,(b) suggests that there are two different dispersion curves, possibly originating from magnetic domains with opposite chirality.

At a twice higher magnetic field of 6~T, the system is in the immediate proximity to the field-driven QCP, where the incommensurate structure gets suppressed due to the collapse of the conical angle and continuously transforms into the field-polarized FM phase. The corresponding INS spectrum is presented in Figs.~\ref{fig:TOF}\,(c,\,g). In contrast to the previous cases, the gap in the spin-wave spectrum closes, and a ring of intensity can be seen in the constant-energy cut at 0.17~meV. Despite the proximity of the system to a collinear field-polarized state, spin fluctuations are still concentrated in the neighborhood of a sphere with radius $\mathbf{q}_\text{h}$, centered at the origin. Two hollow ellipses can be recognized around ($\pm q_\text{h}$\,0~0) and (0~0\,$\pm q_\text{h}$), where the gapless Goldstone modes are now located, and the spin-wave dispersion drops below the level of the constant-energy cut. Despite the magnetic field, which breaks the equivalency of the [100] and [001] axes, the spectrum restores its fourfold symmetry in the $(H\,0\,L)$ plane at the critical field. The momentum-energy cut in Fig.~\ref{fig:TOF}\,(g), parallel to the $q_x$ direction, shows a nearly symmetric parabolic dispersion curve with a vanishing energy gap. There is a small residual difference in the slopes on opposite sides of the dispersion curve, but neither shows the signs of inflection within the available data range, unlike at smaller fields.

Finally, at even higher fields above $B_{\rm c}$, i.e. in the field-polarized FM phase, the spin gap reopens again. This can be well seen in the 8~T data shown in Figs.~\ref{fig:TOF}\,(d,\,h). The value of the spin gap at this field is above the 0.14--0.20~meV integration range of the constant-energy cut, hence no spin-wave intensity can be seen in Figs.~\ref{fig:TOF}\,(d). The magnon dispersion in the momentum-energy cut in Fig.~\ref{fig:TOF}\,(h) is shaped as a symmetric parabola with an energy gap of 0.25~meV.

\begin{figure}[b]
\centerline{\includegraphics[width=\linewidth]{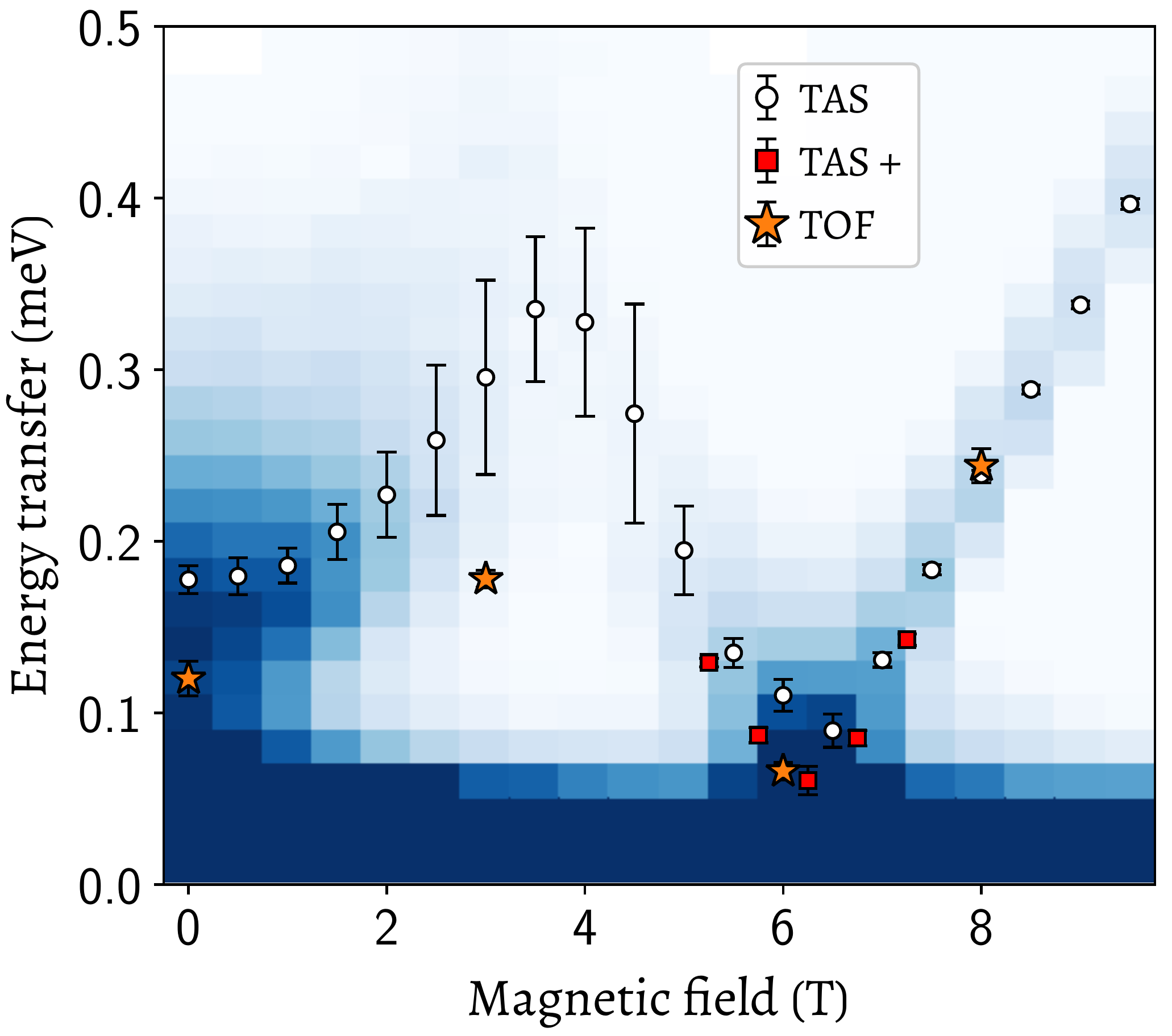}}
\caption{Magnetic-field dependence of the INS spectrum at ($q_\text{h}$\,0\,0), measured as a function of magnetic field applied along the [001] direction. The color map shows INS intensity measured on a TAS instrument (dark-blue color corresponds to high intensity, white to the background level). Open circles show fitted positions of the peak, resulting from Gaussian fits. Red squares show fits of additional TAS data, taken with tighter collimation to improve the resolution (the corresponding raw data are not shown). Orange star symbols correspond to the spin-gap values at the local minimum of the dispersion, extracted from the TOF data.}
\label{fig:TAS}
\end{figure}

To follow the field dependence of the excitation spectrum in ZnCr$_2$Se$_4$ along the $q_x$ direction, we performed an additional TAS experiment, preserving the same experimental geometry. The measurements were taken at a single point ($q_\text{h}$\,0\,0) in momentum space for a number of magnetic fields between 0 and 9.5~T. The results are shown in Fig.~\ref{fig:TAS} in the form of a color map. Here, the horizontal axis corresponds to the magnetic field strength, and the vertical one represents the neutron energy transfer $E$. The color indicates raw INS intensity, with higher intensity shown in dark blue. The broad region of oversaturated intensity near $E=0$ comes from the incoherent elastic line. Every spectrum has been fitted with a Gaussian profile, resulting in the peak positions shown with open circles. In addition, several intermediate fields around the 6~T transition were measured with tighter collimation to improve the resolution. The corresponding fitted peak positions are shown with red squares. Finally, values of the spin gap extracted from TOF data at the local minimum of the dispersion are shown for comparison with orange star symbols.

\begin{figure*}[t!]
\centerline{\includegraphics[width=0.75\linewidth]{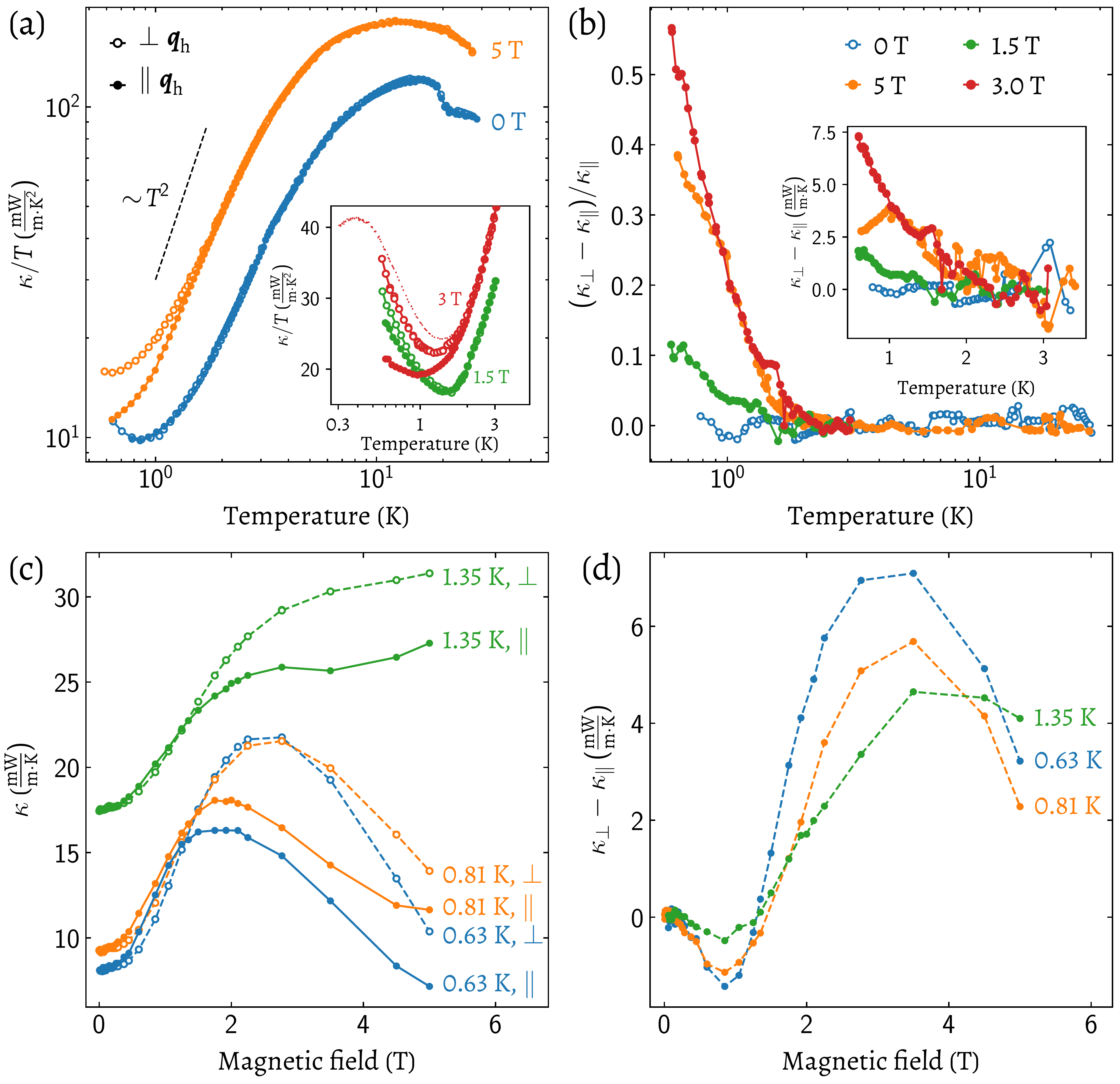}}
\caption{Anisotropy of the thermal conductivity with respect to the direction of the magnetic ordering vector. (a)~Temperature dependence of $\kappa_\perp/T$~($\circ$) and $\kappa_\parallel/T$~($\bullet$) at $B=0$ and 5~T (main panel) and at two intermediate fields of 1.5 and 3~T (inset). The dotted line in the inset shows the 3~T dataset reproduced from Ref.~\cite{GuZhao18}, which we recalculated into $\kappa(T)/T$ and normalized by a constant factor to match with our data in the high-temperature region. (b)~Temperature dependence of the relative thermal-conductivity anisotropy, $(\kappa_{\perp\!}-\kappa_\parallel)/\kappa_\parallel$, at different magnetic fields. The inset shows the low-$T$ part of the same data without normalization, i.e. $\kappa_{\perp\!}-\kappa_\parallel$, on a linear temperature scale. (c)~Magnetic-field dependence of $\kappa_{\perp\!}$~($\circ$) and $\kappa_\parallel$~($\bullet$) for three different temperatures. (d)~Magnetic-field dependence of the thermal-conductivity anisotropy, $\kappa_{\perp\!}-\kappa_\parallel$, resulting from the pairwise subtraction of the data in panel (c).\vspace{-1pt}}
\label{Fig:kappa}
\end{figure*}

First of all, one notices the nonmonotonic behavior of the spin-wave energy at the ($q_\text{h}$\,0\,0) wave vector. It starts from the pseudo-Goldstone magnon gap energy in zero field, then increases to approximately 0.3~meV at intermediate fields and drops down to zero at the critical field, where the transition to the field-polarized FM phase takes place. After that, the spin-wave spectrum is gapped again, and the magnon energy increases linearly with a slope given by the Zeeman energy ($g$-factor). It should be emphasized that in the intermediate field range, within the conical spin-spiral phase, the magnon energy at ($q_\text{h}$\,0\,0) exceeds the actual spin gap energy, because the minima in the magnon dispersion are shifted above and below the scattering plane and cannot be reached in the TAS configuration. This splitting is the main reason for the maximum in the spin-wave energy (open circles) around 3--4~T, which lies above the actual spin-gap energy extracted from the TOF data at the out-of-plane minima of the dispersion (star symbol at 3~T).

The small discrepancy among the TOF and TAS datasets is explained by different momentum resolution. The TAS data without collimation were collected in the double-focusing mode of monochromator and analyzer, which increases the intensity at the expense of the resolution. As a result, the intensity is integrated in a broader vicinity of the dispersion minimum, shifting the center of mass of the peak to somewhat higher energies. Vertical focusing mode with additional collimation reduces this effect, leading to a better agreement with the TOF data.

To summarize, our INS results demonstrate a surprisingly complex non-monotonic behavior of the low-energy magnon spectrum as a function of magnetic field. The previously reported pseudo-Goldstone mode~\cite{TymoshenkoOnykiienko17}, located at the \mbox{$(\pm q_\text{h}\,0~0)$} and \mbox{$(0\,\pm\!q_\text{h}\,0)$} wave vectors in zero field, splits into two separate minima along the $q_z$ direction, and the corresponding spin gap is initially increased. Above 3--4~T, this tendency reverses, and the two dispersion minima merge into a single gapless mode at the critical field, corresponding to the quantum-critical transition to the field-polarized FM state. The collapse of the conical angle of the spin spiral to zero at $B_\text{c}\approx6.25$~T restores the cubic symmetry of the spin-wave spectrum, in spite of the external magnetic field that explicitly breaks this symmetry. At even higher magnetic fields above $B_\text{c}$, the spin-wave spectrum gets fully gapped, the spin gap increasing linearly with the deviation from the quantum critical point.

\vspace{-3pt}\section{Anisotropic thermal conductivity}

The nonmonotonic field dependence of pseudo-Goldstone magnon modes are expected to have pronounced signatures in the magnetic contributions to the specific heat and low-temperature thermal transport properties of ZnCr$_2$Se$_4$. Indeed, the spin-gap energy in zero magnetic field that results from fitting an exponent, $A\exp(-\Delta_\text{PG}/k_\text{B}T)$, to the specific-heat data of Gu \textit{et al.}~\cite{GuZhao18} below 3~K is $\Delta_\text{PG}=0.186$~meV, which perfectly matches the pseudo-Goldstone magnon gap observed in our INS experiments. Note that the actual spin gap in the system (at the ordering wave vector) is expected to be below this value~\cite{TymoshenkoOnykiienko17, ZajdelLi17}, but because the true Goldstone mode carries only 1/3 of the magnon spectral weight, thermodynamic properties are more sensitive to the pseudo-Goldstone magnon gap. With increasing magnetic field, the specific-heat anomaly gradually disappears, resulting in a $\propto T^2$ behavior at the critical field~\cite{GuZhao18}. It is natural to associate this behavior with the closing of the magnon gap observed in our INS measurements. At a higher field of 10~T, the specific-heat anomaly reappears at about 3.5~K, which is in reasonable agreement with the measured spin-gap energy at this field.

Similar anomalies that resemble the nonmonotonic behavior of the spin gap were also observed in the low-temperature thermal conductivity of ZnCr$_2$Se$_4$ as a function of magnetic field and temperature~\cite{GuZhao18}. The zero-field thermal conductivity, $\kappa(T)$, shows a change in behavior around 1~K, which disappears at the critical magnetic field. The corresponding field dependence, $\kappa(B)$, exhibits a nonmonotonic behavior at temperatures below 1~K, with two clear anomalies at the domain-selection field and at the transition to the field-polarized FM phase. In the thermal conductivity experiment from Ref.~\cite{GuZhao18}, the magnetic field was applied along one of the $\langle 111 \rangle$ crystallographic axes, while thermal current was measured in an orthogonal direction~\cite{XFSunPrivate}. In this geometry, the angle formed by the propagation vectors of all three magnetic domains with the magnetic field is approximately the same, hence the distribution of magnetic domains cannot be well controlled (it would be determined only by small unavoidable misalignments of the field direction). It is therefore unclear whether the thermal conductivity was measured in a single- or multidomain state, and how the thermal current was oriented with respect to the magnetic propagation vector.

The total thermal conductivity for a magnetic insulator can be represented as a sum of the phononic (lattice) and magnonic contributions, $\kappa=\kappa_{\rm L}+\kappa_{\rm m}$. The directional dependence of $\kappa$, determined by the thermal gradient $\Delta T$, can yield additional information beyond that afforded by specific heat. An elegant way to separate magnetic effects from those arising entirely from the phonon dispersion is to examine the difference in $\kappa$ measured with $\Delta T$ parallel ($\kappa_\parallel$) and perpendicular ($\kappa_\perp$) to the magnetic ordering vector $\mathbf{q}_{\rm h}$. As seen in section~\ref{Sec:Anisotropy}, the deviations from cubic symmetry in both structural parameters and acoustic phonon frequencies in the magnetically ordered state of ZnCr$_2$Se$_4$ do not exceed 0.1\%. Thus the difference, $\Delta \kappa = \kappa_\perp-\kappa_\parallel$, should be determined exclusively by magnetic effects, which can include magnon heat conduction and anisotropic phonon scattering on magnons and magnetic domain walls.

We have measured anisotropic thermal conductivity in the single-domain state of ZnCr$_2$Se$_4$ in the directions parallel and orthogonal to the magnetic ordering vector in magnetic fields up to 5~T and down to $T \approx 0.6$~K. Figure~\ref{Fig:kappa}\,(a) shows the temperature dependence of $\kappa_\perp/T$ and $\kappa_\parallel/T$ at $B=0$ and 5~T (main panel) and at two intermediate fields of 1.5 and 3~T (inset). No anisotropy is detected in the zero-field measurement, but it can be clearly seen below 2~K in an applied magnetic field. It is more clearly represented by the relative difference, $\Delta\kappa/\kappa \equiv (\kappa_\perp - \kappa_\parallel)/\kappa_\parallel$, which is plotted in Fig.~\ref{Fig:kappa}\,(b). One can see that $\Delta\kappa/\kappa$ reaches above 50\% at the lowest measured temperature, which exceeds the tetragonal distortion by three orders of magnitude and is 250 times higher compared to the anisotropy in the sound velocity at base temperature. Clearly, such a huge effect must have magnetic origin. Remarkably, the anisotropy disappears rapidly with increasing temperature and can no longer be recognized above 2~K, which is still 10 times below the magnetic ordering temperature. To demonstrate that this rapid decrease is not an artefact of normalization to $\kappa_\parallel(T)$, which increases rapidly with temperature, in the inset to Fig.~\ref{Fig:kappa}\,(b) we show the low-temperature part of $\Delta\kappa(T)$ without normalization. Despite the increasing noise towards higher temperatures, the downward trend in the data can be still recognized, and above 2~K the $\Delta\kappa/\kappa$ becomes indistinguishable from zero within the accuracy of our measurements. Neither dilatometry nor ultrasound-velocity measurements presented in section~\ref{Sec:Anisotropy} show anomalies around this temperature, which practically excludes any symmetry-lowering phase transition as a possible origin of the anisotropy onset below 2~K. On the other hand, this temperature, converted to energy units, coincides with the pseudo-Goldstone spin-gap energy, suggesting that a thermally activated process involving low-energy magnons could be responsible.

We have also followed the magnetic-field dependence of $\kappa_\parallel$ and $\kappa_\perp$ at several temperatures, as shown in Fig.~\ref{Fig:kappa}\,(c). The field dependence is nonmonotonic, showing a pronounced maximum around 2 or 2.5~T that exceeds the zero-field value by a factor of 2.0 or 2.7 in the lowest-temperature data for $\kappa_\parallel(B)$ and $\kappa_\perp(B)$, respectively. This non-monotonic behavior is reminiscent of the field dependence of the spin gap discussed in section~\ref{Sec:INS}. In Fig.~\ref{Fig:kappa}\,(d), we show the difference, $\Delta\kappa = \kappa_\perp(B)-\kappa_\parallel(B)$, for each temperature. It also exhibits non-monotonic behavior with a change in sign around 1.5~T for all three temperatures. Although our data are limited to 5~T, an extrapolation to higher fields suggests that the anisotropy should vanish at the critical field of approximately 6~T, as expected from the restored cubic symmetry of the magnon spectrum.

The field- and temperature-dependent data in Figs.~\ref{Fig:kappa}\,(c) and \ref{Fig:kappa}\,(a) reveal that the pronounced low-temperature peak in $\kappa(B)$ with a maximum at $B \approx 2$--2.5~T and the associated maximum in the anisotropy at $B \approx 3~T$ [\ref{Fig:kappa}\,(d)] are related to the low-temperature upturn in $\kappa(T)/T$, which is most pronounced in the 1.5 and 3~T datasets in the inset to Fig.~\ref{Fig:kappa}\,(a). A comparison with the 3 and 5~T datasets published by Gu \textit{et al.}~\cite{GuZhao18}, which extend to somewhat lower temperatures, reveals that $\kappa(T)/T$ has a tendency to decrease again below a certain temperature, and therefore the onset in our data must correspond to the right-hand side of a broad peak in $\kappa(T)/T$ with a maximum at approximately 0.4~K. This is well seen in the inset to Fig.~\ref{Fig:kappa}\,(a), where we reproduce the 3~T dataset of Gu \textit{et al.} for comparison, after normalizing it so that it matches our data at higher temperatures. In spite of the different orientations of the field and thermal currents in the two experiments and a likely difference in the magnetic domain distribution, which may explain a mismatch in the absolute values of the thermal conductivity coefficients, the temperature dependence in both datasets is in qualitative agreement, revealing a pronounced maximum in $\kappa(T)/T$ around 0.4~K. This maximum is most pronounced in the transverse thermal conductivity, $\nabla T\!\perp\mathbf{B}$, and only in the intermediate field range around $B_\text{c}/2$, which enhances $\kappa_\perp(0.5\,\text{K})$ at least sixfold (according to Ref.~\cite{GuZhao18}) compared to its zero-field value. In higher magnetic fields above the field-induced critical point, the monotonic behavior of $\kappa(T)/T$ is restored~\cite{GuZhao18}.

\vspace{-1pt}\section{Discussion and conclusions}

Separating the magnonic ($\kappa_{\rm m}$) and lattice ($\kappa_{\rm L}$) contributions to the thermal conductivity is a challenge given its complex temperature and field dependencies. It is useful to compare $\kappa$ for ZnCr$_2$Se$_4$ to that of the structurally similar helimagnet Cu$_2$OSeO$_3$, which comprises an approximately fcc lattice of Cu$_4$ tetrahedra \cite{PortnichenkoRomhanyi16}. Cu$_2$OSeO$_3$ has $\kappa \approx \text{300--400}$~W/mK at 5~K [while $\kappa_{\rm L} \approx \text{(5--10)}\kappa_{\rm m}$], three orders of magnitude larger than that of ZnCr$_2$Se$_4$ \cite{PrasaiTrump17}. This dramatic difference is clearly related to the stronger coupling between the spin and lattice systems in ZnCr$_2$Se$_4$, as manifested, e.g., in the giant magnetostriction~\cite{HembergerNidda07, FeleaYasin12} (see also section~\ref{Sec:Anisotropy}). Thus both $\kappa_{\rm m}$ and $\kappa_{\rm L}$ in ZnCr$_2$Se$_4$ can be expected to be approximately three orders of magnitude smaller than in Cu$_2$OSeO$_3$, though their relative magnitudes may differ.

We first consider $\kappa_{\rm L}$ and make the reasonable assumption that it is limited at $T\lesssim 5$~K by one-phonon, two-magnon scattering. Such processes, in which a phonon decays into two magnons (and vice versa), are predominated by transverse phonons in this temperature regime having typical velocities in the range 1--2~km/s. The scattering rate $\tau_{\rm ph-m}^{-1}$ is proportional to $B^2q$, where $B$ is an average magnetoelastic (ME) constant and $q$ the phonon wave number \cite{StreibVidalSilva19}. Calculations of $\kappa_{\rm L}$ for ZnCr$_2$Se$_4$ (see Figs.~S1, S2 in the Supplemental Material~\cite{SM}) provide an estimate, $B\simeq 300$~meV, needed to produce a $\kappa_{\rm L}$ equal to the measured $\kappa$ at 5K. For comparison, this value is approximately 50 times larger than the average value for yttrium-iron garnet (YIG) \cite{StreibVidalSilva19}, consistent with very strong spin-phonon coupling in ZnCr$_2$Se$_4$.

The temperature dependence of this scattering is dictated by the magnon occupation factors \cite{StreibVidalSilva19, SM}, $\tau_{\rm ph-m}^{-1}\sim \left(1+n_{\mathbf{q-k}}+n_{\mathbf{k}}\right)$, where $n_{\mathbf{k}}$ is the Bose-Einstein distribution function for a magnon state $\mathbf{k}$. Thus the scattering rate declines with decreasing temperature even for $T\lesssim 2$~K because magnon states satisfying momentum and energy conservation continue to be depopulated. This reduction in scattering can produce an apparent increase of $\kappa_{\rm L}$ relative to the constant scattering rate behavior $\kappa_{\rm L}\propto T^3$ expected for an insulator in the low-$T$ limit. As $T\to 0$, $\tau_{\rm ph-m}^{-1}$ becomes smaller than the boundary scattering rate, leading to $\kappa_{\rm L}\propto T^3$. Thus the overall behavior appears as a plateau or shoulder in $\kappa_{\rm L}(T)$ and a maximum in $\kappa/T$ (Fig.~S2).

\begin{figure}[t]
\centerline{\includegraphics[width=\linewidth]{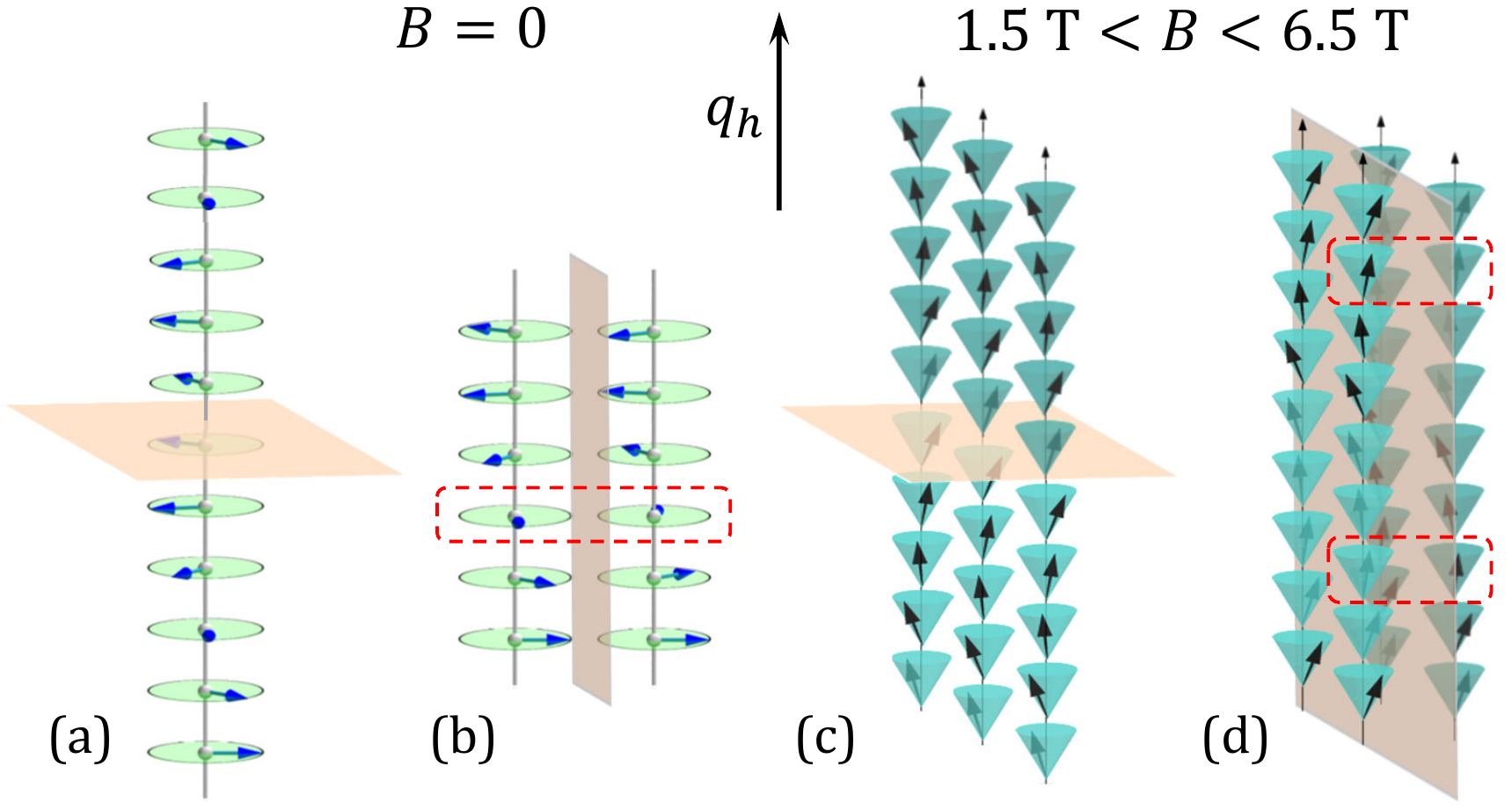}}
\caption{Schematic of planar boundaries between helimagnetic domains of opposite chirality: (a,\,b)~at $B=0$ and (c,\,d)~within the conical phase. The red dashed boxes in (b) and (d) highlight energetically disfavored maximal deviations from FM nearest-neighbor spin orientations.}
\label{Fig:DWalls}
\end{figure}

Though the appearance of an upturn and maximum in $\kappa/T$ [inset, Fig.~\ref{Fig:kappa}\,(a)] may be at least partially associated with $\kappa_{\rm L}$, the anisotropy in $\kappa$ and its nonmonotonic behavior with field at $T\lesssim 2$~K are unlikely to be. The cubic symmetry dictates isotropy in the phonon dispersion. The isotropy of $\kappa$ at $T\gtrsim 2$~K and absence, as noted above, of any identifiable symmetry change in the lattice below 2~K from dilatometry and ultrasonic measurements imply isotropic ME constants as well. To assess potential anisotropy in the phonon scattering, we computed~\cite{SM} the phonon-magnon scattering rate assuming isotropic ME constants and using approximate analytical forms for the three-dimensional magnon dispersion (Fig.~S1, S2)~\cite{TymoshenkoOnykiienko17} constrained by the neutron scattering results along the main symmetry directions\,---\,negligible anisotropy was found. The calculations also yield $\kappa_{\rm L}(B)$ (Fig.~S2) incompatible with the nonmonotonic behavior exhibited in Fig.~\ref{Fig:kappa}\~(c). Gu \textit{et al.}~\cite{GuZhao18} proposed spin fluctuations as the primary scatterers of phonons, but it is unclear how such scattering could produce the nonmonotonic behavior of $\kappa_{\rm L}$ in applied field.

On the other hand, the unusual magnon dispersion and its evolution with applied field described in section~\ref{Sec:INS} suggest that magnon heat conduction is a plausible source for both the anisotropy and nonmonotonic field dependence of $\kappa(B,T)$ [Fig.~\ref{Fig:kappa}\,(c,\,d) and Ref.~\onlinecite{GuZhao18}]. The calculated $\kappa_{\rm m}(B,T)$~\cite{SM}, assuming a constant scattering time (Fig.~S3), captures some of these key features, including $\kappa_{\rm m,\bot}>\kappa_{\rm m,\|}$ and nonmonotonic $\kappa_{\rm m}(B)$ at low $T$. Not surprisingly, these features arise primarily from low-energy magnon states with momenta at small angles to $\mathbf{q}_\text{h}$. Notably, a maximum in $\kappa_{\rm m,\bot}(B)$ at $B=3$~T, consistent with the experiment, occurs due to a substantially larger modulation of the magnon energy at this field for rotations in the plane perpendicular to $\mathbf{q}_\text{h}$~(Fig.~S4)\,---\,a feature that may be related to the splitting of the pseudo-Goldstone mode into two separate modes along the $q_z$ direction [Fig.~5 and Fig.~S1\,(b,\,d)].

While this correspondence between experiment and the calculated $\kappa_{\rm m}$ is encouraging, two discrepancies remain: (i)~the absence of detectable anisotropy in experiment at $B=0$; (ii)~the emergence of anisotropy in experiment only below $T \sim 2$~K. The former is true in spite of the single-domain character of the specimens cooled in a 3~T field to the base temperature ($\sim 0.5$~K) prior to measurements at $B=0$ with increasing $T$. Possible domain relaxation over the time scale of the $\kappa$ measurements up to 1~K (\mbox{$<1$~h}) appears insufficient to explain this discrepancy, as this relaxation only takes place at much higher temperatures (see section~\ref{Sec:Relax}).

The constant magnon scattering time needed to produce a calculated $\kappa_{\rm m}(\text{1~K})$ equal to experimental values ($\kappa\simeq 10^{-2}$~W/mK) is $\tau_{\rm m}\approx 10^{-10}$~s~\cite{SM}. This corresponds to an average mean-free path $\ell_{\rm m}=\langle |v_{\rm m}|\rangle\tau_{\rm m} \simeq 70$~nm, where $\langle |v_{\rm m}|\rangle\simeq 700$~m/s is the magnon velocity averaged over the thermal distribution at \mbox{$T=1$~K}. This value, $\ell_{\rm m}\approx 30\lambda_{\rm h}$ ($\lambda_{\rm h}\equiv 2\pi/q_{\rm h}\approx 2.2$~nm), is substantially shorter than the estimated scattering length for magnon-phonon scattering, and thus another scattering mechanism must limit $\kappa_{\rm m}$, e.g. magnetic disorder. Could this scattering explain both the absence of anisotropy at $B=0$ and the onset of anisotropy in applied field only below $T\simeq 2$~K?

Intriguing candidates for the source of magnon scattering on the appropriate length scale are chiral domain boundaries that coexist within the magnetic single-domain specimen as described in section~\ref{Sec:Relax}. A schematic of possible domain walls separating domains with the same $\mathbf{q}_{\rm h}$ direction but opposite chirality is depicted in Fig.~\ref{Fig:DWalls} for the helical ($B=0$) and conical ($1.5\ {\rm T}\lesssim B\lesssim 6.5\ {\rm T}$) phases. The dominant FM interaction between nearest-neighbor spins makes domain-wall orientations parallel to $\mathbf{q}_{\rm h}$ in the planar helical phase [Fig.~\ref{Fig:DWalls}\,(b)] energetically disfavored, given the periodic modulation of the neighboring spin orientations along the wall that inevitably entail AFM arrangements (dashed area). The neighboring spin orientations in the conical phase along a wall like that of Fig.~\ref{Fig:DWalls}\,(d) are also modulated, with maximal deviation from a FM orientation equal to the cone angle, which decreases with increasing magnetic field. According to these considerations, domain walls are expected to have an anisotropic distribution with the preferred orientation orthogonal to the magnetic ordering vector. For such a distribution, the absence of anisotropy in $\kappa$ at $B=0$ might be explained if heat-carrying magnons scatter from domain walls transverse to their motion more strongly in the conical phase. Within an alternative scenario, in which magnon heat transport is increased along the walls, one might envision unknown dynamical properties of the domain-wall spins on the timescale of $\tau_{\rm m}$ that could be responsible for enhancing $\kappa_{\rm m}$ in the conical phase at $T\lesssim 2$~K. This speculation provides motivation for further investigations where the chiral domain distribution might be manipulated (for example, by electric fields). The properties of domain walls in spiral magnets are not well known and are a topic of broader interest, including for potential applications.

In summary, our studies reveal how the magnetoelastic properties and novel spin-wave spectrum of ZnCr$_2$Se$_4$ evolve in applied magnetic field within its spin-spiral phase and identify the anisotropic thermal conductivity to be a sensitive probe of the low-energy magnon spectrum. The nonmonotonic field dependency of the latter at $T\lesssim 2$~K strongly favors an interpretation in which transport by magnons is principally responsible. The possible role of magnon scattering by or transport along two-dimensional domain walls associated with different spin-helix chiralities is proposed as an interesting issue for future investigation.

\vspace{-1pt}\section*{Acknowledgments}

We would like to thank C.~Hess, S.-C. Lee, D.~Efremov, A.~Yaresko, and M.~Vojta for stimulating discussions and X.-F. Sun for sharing their recently published thermal-conductivity data~\cite{GuZhao18}. This project was funded by the German Research Foundation (DFG) through the Collaborative Research Center SFB~1143 in Dresden [projects C03 (D.\,S.\,I.) and A06 (S.\,R.)], individual research grant IN\,\mbox{209/4-1}, and the Trans\-regional Collaborative Research Center TRR~80 (Augsburg, Munich, Stuttgart). T.~M. and R.~T. \mbox{acknowledge} support from DFG-SFB 1170 ToCoTronics (project B04) and the ERC Starting Grant ERC-StG-Thomale-336012 ``Topolectrics''. Work at the University of Miami was supported by the U.S. Department of Energy (DOE), Office of Science, Basic Energy Sciences (BES), under Award No.~DE-SC0008607.

\bibliography{ZnCr2Se4}\vspace{-3pt}

\begin{thebibliography}{45}%
\makeatletter
\providecommand \@ifxundefined [1]{%
 \@ifx{#1\undefined}
}%
\providecommand \@ifnum [1]{%
 \ifnum #1\expandafter \@firstoftwo
 \else \expandafter \@secondoftwo
 \fi
}%
\providecommand \@ifx [1]{%
 \ifx #1\expandafter \@firstoftwo
 \else \expandafter \@secondoftwo
 \fi
}%
\providecommand \natexlab [1]{#1}%
\providecommand \enquote  [1]{``#1''}%
\providecommand \bibnamefont  [1]{#1}%
\providecommand \bibfnamefont [1]{#1}%
\providecommand \citenamefont [1]{#1}%
\providecommand \href@noop [0]{\@secondoftwo}%
\providecommand \href [0]{\begingroup \@sanitize@url \@href}%
\providecommand \@href[1]{\@@startlink{#1}\@@href}%
\providecommand \@@href[1]{\endgroup#1\@@endlink}%
\providecommand \@sanitize@url [0]{\catcode `\\12\catcode `\$12\catcode
  `\&12\catcode `\#12\catcode `\^12\catcode `\_12\catcode `\%12\relax}%
\providecommand \@@startlink[1]{}%
\providecommand \@@endlink[0]{}%
\providecommand \url  [0]{\begingroup\@sanitize@url \@url }%
\providecommand \@url [1]{\endgroup\@href {#1}{\urlprefix }}%
\providecommand \urlprefix  [0]{URL }%
\providecommand \Eprint [0]{\href }%
\providecommand \doibase [0]{https://doi.org/}%
\providecommand \selectlanguage [0]{\@gobble}%
\providecommand \bibinfo  [0]{\@secondoftwo}%
\providecommand \bibfield  [0]{\@secondoftwo}%
\providecommand \translation [1]{[#1]}%
\providecommand \BibitemOpen [0]{}%
\providecommand \bibitemStop [0]{}%
\providecommand \bibitemNoStop [0]{.\EOS\space}%
\providecommand \EOS [0]{\spacefactor3000\relax}%
\providecommand \BibitemShut  [1]{\csname bibitem#1\endcsname}%
\let\auto@bib@innerbib\@empty
\bibitem [{\citenamefont {Yaresko}(2008)}]{Yaresko08}%
  \BibitemOpen
  \bibfield  {author} {\bibinfo {author} {\bibfnamefont {A.~N.}\ \bibnamefont
  {Yaresko}},\ }\bibfield  {title} {\bibinfo {title} {\textit{Electronic band
  structure and exchange coupling constants in
  $A{\kern.5pt}\text{Cr}_\text{\!2}X_\text{4}$ spinels ($A$\,\,=~Zn, Cd, Hg;
  $X$\,=~O, S, Se)}},\ }\href {https://doi.org/10.1103/PhysRevB.77.115106}
  {\bibfield  {journal} {\bibinfo  {journal} {Phys. Rev.~B}\ }\textbf {\bibinfo
  {volume} {77}},\ \bibinfo {pages} {115106} (\bibinfo {year}
  {2008})}\BibitemShut {NoStop}%
\bibitem [{\citenamefont {Moessner}\ and\ \citenamefont
  {Chalker}(1998)}]{MoessnerChalker98}%
  \BibitemOpen
  \bibfield  {author} {\bibinfo {author} {\bibfnamefont {R.}~\bibnamefont
  {Moessner}}\ and\ \bibinfo {author} {\bibfnamefont {J.~T.}\ \bibnamefont
  {Chalker}},\ }\bibfield  {title} {\bibinfo {title} {\textit{Properties of a
  classical spin liquid: The Heisenberg pyrochlore antiferromagnet}},\ }\href
  {https://doi.org/10.1103/PhysRevLett.80.2929} {\bibfield  {journal} {\bibinfo
   {journal} {Phys. Rev. Lett.}\ }\textbf {\bibinfo {volume} {80}},\ \bibinfo
  {pages} {2929} (\bibinfo {year} {1998})}\BibitemShut {NoStop}%
\bibitem [{\citenamefont {Benton}\ \emph {et~al.}(2016)\citenamefont {Benton},
  \citenamefont {Jaubert}, \citenamefont {Yan},\ and\ \citenamefont
  {Shannon}}]{BentonJaubert16}%
  \BibitemOpen
  \bibfield  {author} {\bibinfo {author} {\bibfnamefont {O.}~\bibnamefont
  {Benton}}, \bibinfo {author} {\bibfnamefont {L.~D.~C.}\ \bibnamefont
  {Jaubert}}, \bibinfo {author} {\bibfnamefont {H.}~\bibnamefont {Yan}},\ and\
  \bibinfo {author} {\bibfnamefont {N.}~\bibnamefont {Shannon}},\ }\bibfield
  {title} {\bibinfo {title} {\textit{A spin-liquid with pinch-line
  singularities on the pyrochlore lattice}},\ }\href
  {https://doi.org/10.1038/ncomms11572} {\bibfield  {journal} {\bibinfo
  {journal} {Nat. Commun.}\ }\textbf {\bibinfo {volume} {7}},\ \bibinfo {pages}
  {11572} (\bibinfo {year} {2016})}\BibitemShut {NoStop}%
\bibitem [{\citenamefont {Okubo}\ \emph {et~al.}(2011)\citenamefont {Okubo},
  \citenamefont {Nguyen},\ and\ \citenamefont {Kawamura}}]{OkuboNguyen11}%
  \BibitemOpen
  \bibfield  {author} {\bibinfo {author} {\bibfnamefont {T.}~\bibnamefont
  {Okubo}}, \bibinfo {author} {\bibfnamefont {T.~H.}\ \bibnamefont {Nguyen}},\
  and\ \bibinfo {author} {\bibfnamefont {H.}~\bibnamefont {Kawamura}},\
  }\bibfield  {title} {\bibinfo {title} {\textit{Cubic and noncubic
  multiple-$q$ states in the Heisenberg antiferromagnet on the pyrochlore
  lattice}},\ }\href {https://doi.org/10.1103/PhysRevB.84.144432} {\bibfield
  {journal} {\bibinfo  {journal} {Phys. Rev.~B}\ }\textbf {\bibinfo {volume}
  {84}},\ \bibinfo {pages} {144432} (\bibinfo {year} {2011})}\BibitemShut
  {NoStop}%
\bibitem [{\citenamefont {Tymoshenko}\ \emph {et~al.}(2017)\citenamefont
  {Tymoshenko}, \citenamefont {Onykiienko}, \citenamefont {M\"uller},
  \citenamefont {Thomale}, \citenamefont {Rachel}, \citenamefont {Cameron},
  \citenamefont {Portnichenko}, \citenamefont {Efremov}, \citenamefont
  {Tsurkan}, \citenamefont {Abernathy}, \citenamefont {Ollivier}, \citenamefont
  {Schneidewind}, \citenamefont {Piovano}, \citenamefont {Felea}, \citenamefont
  {Loidl},\ and\ \citenamefont {Inosov}}]{TymoshenkoOnykiienko17}%
  \BibitemOpen
  \bibfield  {author} {\bibinfo {author} {\bibfnamefont {Y.~V.}\ \bibnamefont
  {Tymoshenko}}, \bibinfo {author} {\bibfnamefont {Y.~A.}\ \bibnamefont
  {Onykiienko}}, \bibinfo {author} {\bibfnamefont {T.}~\bibnamefont
  {M\"uller}}, \bibinfo {author} {\bibfnamefont {R.}~\bibnamefont {Thomale}},
  \bibinfo {author} {\bibfnamefont {S.}~\bibnamefont {Rachel}}, \bibinfo
  {author} {\bibfnamefont {A.~S.}\ \bibnamefont {Cameron}}, \bibinfo {author}
  {\bibfnamefont {P.~Y.}\ \bibnamefont {Portnichenko}}, \bibinfo {author}
  {\bibfnamefont {D.~V.}\ \bibnamefont {Efremov}}, \bibinfo {author}
  {\bibfnamefont {V.}~\bibnamefont {Tsurkan}}, \bibinfo {author} {\bibfnamefont
  {D.~L.}\ \bibnamefont {Abernathy}}, \bibinfo {author} {\bibfnamefont
  {J.}~\bibnamefont {Ollivier}}, \bibinfo {author} {\bibfnamefont
  {A.}~\bibnamefont {Schneidewind}}, \bibinfo {author} {\bibfnamefont
  {A.}~\bibnamefont {Piovano}}, \bibinfo {author} {\bibfnamefont
  {V.}~\bibnamefont {Felea}}, \bibinfo {author} {\bibfnamefont
  {A.}~\bibnamefont {Loidl}},\ and\ \bibinfo {author} {\bibfnamefont {D.~S.}\
  \bibnamefont {Inosov}},\ }\bibfield  {title} {\bibinfo {title}
  {\textit{Pseudo-Goldstone magnons in the frustrated \mbox{$S=3/2$} Heisenberg
  helimagnet {ZnCr}$_\text{\!2}${Se}$_\text{4}$ with a pyrochlore magnetic
  sublattice}},\ }\href {https://doi.org/10.1103/PhysRevX.7.041049} {\bibfield
  {journal} {\bibinfo  {journal} {Phys. Rev.~X}\ }\textbf {\bibinfo {volume}
  {7}},\ \bibinfo {pages} {041049} (\bibinfo {year} {2017})}\BibitemShut
  {NoStop}%
\bibitem [{\citenamefont {Plumier}(1966)}]{Plumier66}%
  \BibitemOpen
  \bibfield  {author} {\bibinfo {author} {\bibfnamefont {R.~J.}\ \bibnamefont
  {Plumier}},\ }\bibfield  {title} {\bibinfo {title} {\textit{Neutron
  diffraction study of helimagnetic spinel ZnCr$_\text{\!2}$Se$_\text{4}$}},\
  }\href {https://doi.org/10.1063/1.1708540} {\bibfield  {journal} {\bibinfo
  {journal} {J. Appl. Phys.}\ }\textbf {\bibinfo {volume} {37}},\ \bibinfo
  {pages} {964} (\bibinfo {year} {1966})}\BibitemShut {NoStop}%
\bibitem [{\citenamefont {Hidaka}\ \emph
  {et~al.}(2003{\natexlab{a}})\citenamefont {Hidaka}, \citenamefont {Tokiwa},
  \citenamefont {Fujii}, \citenamefont {Watanabe},\ and\ \citenamefont
  {Akimitsu}}]{HidakaTokiwa03}%
  \BibitemOpen
  \bibfield  {author} {\bibinfo {author} {\bibfnamefont {M.}~\bibnamefont
  {Hidaka}}, \bibinfo {author} {\bibfnamefont {N.}~\bibnamefont {Tokiwa}},
  \bibinfo {author} {\bibfnamefont {M.}~\bibnamefont {Fujii}}, \bibinfo
  {author} {\bibfnamefont {S.}~\bibnamefont {Watanabe}},\ and\ \bibinfo
  {author} {\bibfnamefont {J.}~\bibnamefont {Akimitsu}},\ }\bibfield  {title}
  {\bibinfo {title} {\textit{Correlation between the structural and
  antiferromagnetic phase transitions in ZnCr$_\text{\!2}$Se$_\text{4}$}},\
  }\href {https://doi.org/10.1002/pssb.200301502} {\bibfield  {journal}
  {\bibinfo  {journal} {Phys. Stat. Sol. (b)}\ }\textbf {\bibinfo {volume}
  {236}},\ \bibinfo {pages} {9} (\bibinfo {year}
  {2003}{\natexlab{a}})}\BibitemShut {NoStop}%
\bibitem [{\citenamefont {Yokaichiya}\ \emph {et~al.}(2009)\citenamefont
  {Yokaichiya}, \citenamefont {Krimmel}, \citenamefont {Tsurkan}, \citenamefont
  {Margiolaki}, \citenamefont {Thompson}, \citenamefont {Bordallo},
  \citenamefont {Buchsteiner}, \citenamefont {St\"u\ss{}er}, \citenamefont
  {Argyriou},\ and\ \citenamefont {Loidl}}]{YokaichiyaKrimmel09}%
  \BibitemOpen
  \bibfield  {author} {\bibinfo {author} {\bibfnamefont {F.}~\bibnamefont
  {Yokaichiya}}, \bibinfo {author} {\bibfnamefont {A.}~\bibnamefont {Krimmel}},
  \bibinfo {author} {\bibfnamefont {V.}~\bibnamefont {Tsurkan}}, \bibinfo
  {author} {\bibfnamefont {I.}~\bibnamefont {Margiolaki}}, \bibinfo {author}
  {\bibfnamefont {P.}~\bibnamefont {Thompson}}, \bibinfo {author}
  {\bibfnamefont {H.~N.}\ \bibnamefont {Bordallo}}, \bibinfo {author}
  {\bibfnamefont {A.}~\bibnamefont {Buchsteiner}}, \bibinfo {author}
  {\bibfnamefont {N.}~\bibnamefont {St\"u\ss{}er}}, \bibinfo {author}
  {\bibfnamefont {D.~N.}\ \bibnamefont {Argyriou}},\ and\ \bibinfo {author}
  {\bibfnamefont {A.}~\bibnamefont {Loidl}},\ }\bibfield  {title} {\bibinfo
  {title} {\textit{Spin-driven phase transitions in
  ZnCr$_\text{\!2}$Se$_\text{4}$ and ZnCr$_\text{\!2}$S$_\text{4}$ probed by
  high-resolution synchrotron x-ray and neutron powder diffraction}},\ }\href
  {https://doi.org/10.1103/PhysRevB.79.064423} {\bibfield  {journal} {\bibinfo
  {journal} {Phys. Rev.~B}\ }\textbf {\bibinfo {volume} {79}},\ \bibinfo
  {pages} {064423} (\bibinfo {year} {2009})}\BibitemShut {NoStop}%
\bibitem [{\citenamefont {Cameron}\ \emph {et~al.}(2016)\citenamefont
  {Cameron}, \citenamefont {Tymoshenko}, \citenamefont {Portnichenko},
  \citenamefont {Gavilano}, \citenamefont {Tsurkan}, \citenamefont {Felea},
  \citenamefont {Loidl}, \citenamefont {Zherlitsyn}, \citenamefont {Wosnitza},\
  and\ \citenamefont {Inosov}}]{CameronTymoshenko16}%
  \BibitemOpen
  \bibfield  {author} {\bibinfo {author} {\bibfnamefont {A.~S.}\ \bibnamefont
  {Cameron}}, \bibinfo {author} {\bibfnamefont {Y.~V.}\ \bibnamefont
  {Tymoshenko}}, \bibinfo {author} {\bibfnamefont {P.~Y.}\ \bibnamefont
  {Portnichenko}}, \bibinfo {author} {\bibfnamefont {J.}~\bibnamefont
  {Gavilano}}, \bibinfo {author} {\bibfnamefont {V.}~\bibnamefont {Tsurkan}},
  \bibinfo {author} {\bibfnamefont {V.}~\bibnamefont {Felea}}, \bibinfo
  {author} {\bibfnamefont {A.}~\bibnamefont {Loidl}}, \bibinfo {author}
  {\bibfnamefont {S.}~\bibnamefont {Zherlitsyn}}, \bibinfo {author}
  {\bibfnamefont {J.}~\bibnamefont {Wosnitza}},\ and\ \bibinfo {author}
  {\bibfnamefont {D.~S.}\ \bibnamefont {Inosov}},\ }\bibfield  {title}
  {\bibinfo {title} {\textit{Magnetic phase diagram of the helimagnetic spinel
  compound ZnCr$_\text{\!2}$Se$_\text{4}$ revisited by small-angle neutron
  scattering}},\ }\href {https://doi.org/10.1088/0953-8984/28/14/146001}
  {\bibfield  {journal} {\bibinfo  {journal} {J.~Phys.: Condens. Matter}\
  }\textbf {\bibinfo {volume} {28}},\ \bibinfo {pages} {146001} (\bibinfo
  {year} {2016})}\BibitemShut {NoStop}%
\bibitem [{\citenamefont {Baltzer}\ \emph {et~al.}(1965)\citenamefont
  {Baltzer}, \citenamefont {Lehmann},\ and\ \citenamefont
  {Robbins}}]{BaltzerLehmann65}%
  \BibitemOpen
  \bibfield  {author} {\bibinfo {author} {\bibfnamefont {P.~K.}\ \bibnamefont
  {Baltzer}}, \bibinfo {author} {\bibfnamefont {H.~W.}\ \bibnamefont
  {Lehmann}},\ and\ \bibinfo {author} {\bibfnamefont {M.}~\bibnamefont
  {Robbins}},\ }\bibfield  {title} {\bibinfo {title} {\textit{Insulating
  ferromagnetic spinels}},\ }\href {https://doi.org/10.1103/PhysRevLett.15.493}
  {\bibfield  {journal} {\bibinfo  {journal} {Phys. Rev. Lett.}\ }\textbf
  {\bibinfo {volume} {15}},\ \bibinfo {pages} {493} (\bibinfo {year}
  {1965})}\BibitemShut {NoStop}%
\bibitem [{\citenamefont {Hemberger}\ \emph {et~al.}(2007)\citenamefont
  {Hemberger}, \citenamefont {von Nidda}, \citenamefont {Tsurkan},\ and\
  \citenamefont {Loidl}}]{HembergerNidda07}%
  \BibitemOpen
  \bibfield  {author} {\bibinfo {author} {\bibfnamefont {J.}~\bibnamefont
  {Hemberger}}, \bibinfo {author} {\bibfnamefont {H.-A.~K.}\ \bibnamefont {von
  Nidda}}, \bibinfo {author} {\bibfnamefont {V.}~\bibnamefont {Tsurkan}},\ and\
  \bibinfo {author} {\bibfnamefont {A.}~\bibnamefont {Loidl}},\ }\bibfield
  {title} {\bibinfo {title} {\textit{Large magnetostriction and negative
  thermal expansion in the frustrated antiferromagnet
  ZnCr$_\text{\!2}$Se$_\text{4}$}},\ }\href
  {https://doi.org/10.1103/PhysRevLett.98.147203} {\bibfield  {journal}
  {\bibinfo  {journal} {Phys. Rev. Lett.}\ }\textbf {\bibinfo {volume} {98}},\
  \bibinfo {pages} {147203} (\bibinfo {year} {2007})}\BibitemShut {NoStop}%
\bibitem [{\citenamefont {Zajdel}\ \emph {et~al.}(2017)\citenamefont {Zajdel},
  \citenamefont {Li}, \citenamefont {van Beek}, \citenamefont {Lappas},
  \citenamefont {Ziolkowska}, \citenamefont {Jaskiewicz}, \citenamefont
  {Stock},\ and\ \citenamefont {Green}}]{ZajdelLi17}%
  \BibitemOpen
  \bibfield  {author} {\bibinfo {author} {\bibfnamefont {P.}~\bibnamefont
  {Zajdel}}, \bibinfo {author} {\bibfnamefont {W.-Y.}\ \bibnamefont {Li}},
  \bibinfo {author} {\bibfnamefont {W.}~\bibnamefont {van Beek}}, \bibinfo
  {author} {\bibfnamefont {A.}~\bibnamefont {Lappas}}, \bibinfo {author}
  {\bibfnamefont {A.}~\bibnamefont {Ziolkowska}}, \bibinfo {author}
  {\bibfnamefont {S.}~\bibnamefont {Jaskiewicz}}, \bibinfo {author}
  {\bibfnamefont {C.}~\bibnamefont {Stock}},\ and\ \bibinfo {author}
  {\bibfnamefont {M.~A.}\ \bibnamefont {Green}},\ }\bibfield  {title} {\bibinfo
  {title} {\textit{Structure and magnetism in the bond-frustrated spinel
  ZnCr$_\text{\!2}$Se$_\text{4}$}},\ }\href
  {https://doi.org/10.1103/PhysRevB.95.134401} {\bibfield  {journal} {\bibinfo
  {journal} {Phys. Rev.~B}\ }\textbf {\bibinfo {volume} {95}},\ \bibinfo
  {pages} {134401} (\bibinfo {year} {2017})}\BibitemShut {NoStop}%
\bibitem [{\citenamefont {Larkin}\ and\ \citenamefont
  {Pikin}(1969)}]{LarkinPikin69}%
  \BibitemOpen
  \bibfield  {author} {\bibinfo {author} {\bibfnamefont {A.~I.}\ \bibnamefont
  {Larkin}}\ and\ \bibinfo {author} {\bibfnamefont {S.~A.}\ \bibnamefont
  {Pikin}},\ }\bibfield  {title} {\bibinfo {title} {\textit{Phase transitions
  of the first order but nearly of the second}},\ }\href@noop {} {\bibfield
  {journal} {\bibinfo  {journal} {Sov. Phys. JETP}\ }\textbf {\bibinfo {volume}
  {29}},\ \bibinfo {pages} {891} (\bibinfo {year} {1969})}\BibitemShut
  {NoStop}%
\bibitem [{\citenamefont {Kleinberger}\ and\ \citenamefont {{de
  Kouchkovsky}}(1966)}]{KleinbergerKouchkovsky66}%
  \BibitemOpen
  \bibfield  {author} {\bibinfo {author} {\bibfnamefont {R.}~\bibnamefont
  {Kleinberger}}\ and\ \bibinfo {author} {\bibfnamefont {R.}~\bibnamefont {{de
  Kouchkovsky}}},\ }\bibfield  {title} {\bibinfo {title} {\textit{{\'E}tude
  radiocristallographique {\`a} basse temp{\'e}rature du spinelle
  ZnCr$_\text{\!2}$Se$_\text{4}$}},\ }\href@noop {} {\bibfield  {journal}
  {\bibinfo  {journal} {C. R. Acad. Sci. Paris Ser. B}\ }\textbf {\bibinfo
  {volume} {262}},\ \bibinfo {pages} {628} (\bibinfo {year}
  {1966})}\BibitemShut {NoStop}%
\bibitem [{\citenamefont {Hidaka}\ \emph
  {et~al.}(2003{\natexlab{b}})\citenamefont {Hidaka}, \citenamefont
  {Yoshimura}, \citenamefont {Tokiwa}, \citenamefont {Akimitsu}, \citenamefont
  {Park}, \citenamefont {Park}, \citenamefont {Ji},\ and\ \citenamefont
  {Lee}}]{HidakaYoshimura03}%
  \BibitemOpen
  \bibfield  {author} {\bibinfo {author} {\bibfnamefont {M.}~\bibnamefont
  {Hidaka}}, \bibinfo {author} {\bibfnamefont {M.}~\bibnamefont {Yoshimura}},
  \bibinfo {author} {\bibfnamefont {N.}~\bibnamefont {Tokiwa}}, \bibinfo
  {author} {\bibfnamefont {J.}~\bibnamefont {Akimitsu}}, \bibinfo {author}
  {\bibfnamefont {Y.~J.}\ \bibnamefont {Park}}, \bibinfo {author}
  {\bibfnamefont {J.~H.}\ \bibnamefont {Park}}, \bibinfo {author}
  {\bibfnamefont {S.~D.}\ \bibnamefont {Ji}},\ and\ \bibinfo {author}
  {\bibfnamefont {K.~B.}\ \bibnamefont {Lee}},\ }\bibfield  {title} {\bibinfo
  {title} {\textit{Structural modulation inducedby the incommensurate
  antiferromagnetic phase transitionin ZnCr$_\text{2}$Se$_\text{4}$}},\ }\href
  {https://doi.org/10.1002/pssb.200301551} {\bibfield  {journal} {\bibinfo
  {journal} {phys. stat. sol. (b)}\ }\textbf {\bibinfo {volume} {236}},\
  \bibinfo {pages} {570} (\bibinfo {year} {2003}{\natexlab{b}})}\BibitemShut
  {NoStop}%
\bibitem [{\citenamefont {Chen}\ \emph {et~al.}(2014)\citenamefont {Chen},
  \citenamefont {Yang}, \citenamefont {Tong}, \citenamefont {Huang},
  \citenamefont {Zhang}, \citenamefont {Zhang}, \citenamefont {Song},
  \citenamefont {Pi}, \citenamefont {Sun}, \citenamefont {Tian},\ and\
  \citenamefont {Zhang}}]{ChenYang14}%
  \BibitemOpen
  \bibfield  {author} {\bibinfo {author} {\bibfnamefont {X.~L.}\ \bibnamefont
  {Chen}}, \bibinfo {author} {\bibfnamefont {Z.~R.}\ \bibnamefont {Yang}},
  \bibinfo {author} {\bibfnamefont {W.}~\bibnamefont {Tong}}, \bibinfo {author}
  {\bibfnamefont {Z.~H.}\ \bibnamefont {Huang}}, \bibinfo {author}
  {\bibfnamefont {L.}~\bibnamefont {Zhang}}, \bibinfo {author} {\bibfnamefont
  {S.~L.}\ \bibnamefont {Zhang}}, \bibinfo {author} {\bibfnamefont {W.~H.}\
  \bibnamefont {Song}}, \bibinfo {author} {\bibfnamefont {L.}~\bibnamefont
  {Pi}}, \bibinfo {author} {\bibfnamefont {Y.~P.}\ \bibnamefont {Sun}},
  \bibinfo {author} {\bibfnamefont {M.~L.}\ \bibnamefont {Tian}},\ and\
  \bibinfo {author} {\bibfnamefont {Y.~H.}\ \bibnamefont {Zhang}},\ }\bibfield
  {title} {\bibinfo {title} {\textit{Study of negative thermal expansion in the
  frustrated spinel ZnCr$_\text{2}$Se$_\text{4}$}},\ }\href
  {https://doi.org/10.1063/1.4867217} {\bibfield  {journal} {\bibinfo
  {journal} {J.~Appl. Phys.}\ }\textbf {\bibinfo {volume} {115}},\ \bibinfo
  {pages} {083916} (\bibinfo {year} {2014})}\BibitemShut {NoStop}%
\bibitem [{\citenamefont {Rudolf}\ \emph
  {et~al.}(2007{\natexlab{a}})\citenamefont {Rudolf}, \citenamefont {Kant},
  \citenamefont {Mayr}, \citenamefont {Hemberger}, \citenamefont {Tsurkan},\
  and\ \citenamefont {Loidl}}]{RudolfKant07prb}%
  \BibitemOpen
  \bibfield  {author} {\bibinfo {author} {\bibfnamefont {T.}~\bibnamefont
  {Rudolf}}, \bibinfo {author} {\bibfnamefont {C.}~\bibnamefont {Kant}},
  \bibinfo {author} {\bibfnamefont {F.}~\bibnamefont {Mayr}}, \bibinfo {author}
  {\bibfnamefont {J.}~\bibnamefont {Hemberger}}, \bibinfo {author}
  {\bibfnamefont {V.}~\bibnamefont {Tsurkan}},\ and\ \bibinfo {author}
  {\bibfnamefont {A.}~\bibnamefont {Loidl}},\ }\bibfield  {title} {\bibinfo
  {title} {\textit{Spin-phonon coupling in
  {Z}n{C}r$_\text{\!2}${S}e$_\text{4}$}},\ }\href
  {https://doi.org/10.1103/PhysRevB.75.052410} {\bibfield  {journal} {\bibinfo
  {journal} {Phys. Rev.~B}\ }\textbf {\bibinfo {volume} {75}},\ \bibinfo
  {pages} {052410} (\bibinfo {year} {2007}{\natexlab{a}})}\BibitemShut
  {NoStop}%
\bibitem [{\citenamefont {Rudolf}\ \emph
  {et~al.}(2007{\natexlab{b}})\citenamefont {Rudolf}, \citenamefont {Kant},
  \citenamefont {Mayr}, \citenamefont {Hemberger}, \citenamefont {Tsurkan},\
  and\ \citenamefont {Loidl}}]{RudolfKant07njp}%
  \BibitemOpen
  \bibfield  {author} {\bibinfo {author} {\bibfnamefont {T.}~\bibnamefont
  {Rudolf}}, \bibinfo {author} {\bibfnamefont {C.}~\bibnamefont {Kant}},
  \bibinfo {author} {\bibfnamefont {F.}~\bibnamefont {Mayr}}, \bibinfo {author}
  {\bibfnamefont {J.}~\bibnamefont {Hemberger}}, \bibinfo {author}
  {\bibfnamefont {V.}~\bibnamefont {Tsurkan}},\ and\ \bibinfo {author}
  {\bibfnamefont {A.}~\bibnamefont {Loidl}},\ }\bibfield  {title} {\bibinfo
  {title} {\textit{Spin-phonon coupling in antiferromagnetic chromium
  spinels}},\ }\href {https://doi.org/10.1088/1367-2630/9/3/076} {\bibfield
  {journal} {\bibinfo  {journal} {New J.~Phys.}\ }\textbf {\bibinfo {volume}
  {9}},\ \bibinfo {pages} {76} (\bibinfo {year}
  {2007}{\natexlab{b}})}\BibitemShut {NoStop}%
\bibitem [{\citenamefont {Felea}\ \emph {et~al.}(2012)\citenamefont {Felea},
  \citenamefont {Yasin}, \citenamefont {G\"unther}, \citenamefont
  {Deisenhofer}, \citenamefont {Krug~von Nidda}, \citenamefont {Zherlitsyn},
  \citenamefont {Tsurkan}, \citenamefont {Lemmens}, \citenamefont {Wosnitza},\
  and\ \citenamefont {Loidl}}]{FeleaYasin12}%
  \BibitemOpen
  \bibfield  {author} {\bibinfo {author} {\bibfnamefont {V.}~\bibnamefont
  {Felea}}, \bibinfo {author} {\bibfnamefont {S.}~\bibnamefont {Yasin}},
  \bibinfo {author} {\bibfnamefont {A.}~\bibnamefont {G\"unther}}, \bibinfo
  {author} {\bibfnamefont {J.}~\bibnamefont {Deisenhofer}}, \bibinfo {author}
  {\bibfnamefont {H.-A.}\ \bibnamefont {Krug~von Nidda}}, \bibinfo {author}
  {\bibfnamefont {S.}~\bibnamefont {Zherlitsyn}}, \bibinfo {author}
  {\bibfnamefont {V.}~\bibnamefont {Tsurkan}}, \bibinfo {author} {\bibfnamefont
  {P.}~\bibnamefont {Lemmens}}, \bibinfo {author} {\bibfnamefont
  {J.}~\bibnamefont {Wosnitza}},\ and\ \bibinfo {author} {\bibfnamefont
  {A.}~\bibnamefont {Loidl}},\ }\bibfield  {title} {\bibinfo {title}
  {\textit{Spin-lattice coupling in the frustrated antiferromagnet
  ZnCr$_\text{\!2}$Se$_\text{4}$ probed by ultrasound}},\ }\href
  {https://doi.org/10.1103/PhysRevB.86.104420} {\bibfield  {journal} {\bibinfo
  {journal} {Phys. Rev.~B}\ }\textbf {\bibinfo {volume} {86}},\ \bibinfo
  {pages} {104420} (\bibinfo {year} {2012})}\BibitemShut {NoStop}%
\bibitem [{\citenamefont {Murakawa}\ \emph {et~al.}(2008)\citenamefont
  {Murakawa}, \citenamefont {Onose}, \citenamefont {Ohgushi}, \citenamefont
  {Ishiwata},\ and\ \citenamefont {Tokura}}]{MurakawaOnose08}%
  \BibitemOpen
  \bibfield  {author} {\bibinfo {author} {\bibfnamefont {H.}~\bibnamefont
  {Murakawa}}, \bibinfo {author} {\bibfnamefont {Y.}~\bibnamefont {Onose}},
  \bibinfo {author} {\bibfnamefont {K.}~\bibnamefont {Ohgushi}}, \bibinfo
  {author} {\bibfnamefont {S.}~\bibnamefont {Ishiwata}},\ and\ \bibinfo
  {author} {\bibfnamefont {Y.}~\bibnamefont {Tokura}},\ }\bibfield  {title}
  {\bibinfo {title} {\textit{Generation of electric polarization with rotating
  magnetic field in helimagnet ZnCr$_\text{2}$Se$_\text{4}$}},\ }\href
  {https://doi.org/10.1143/JPSJ.77.043709} {\bibfield  {journal} {\bibinfo
  {journal} {J.~Phys. Soc. Jpn.}\ }\textbf {\bibinfo {volume} {77}},\ \bibinfo
  {pages} {043709} (\bibinfo {year} {2008})}\BibitemShut {NoStop}%
\bibitem [{\citenamefont {Tokura}\ and\ \citenamefont
  {Seki}(2010)}]{TokuraSeki10}%
  \BibitemOpen
  \bibfield  {author} {\bibinfo {author} {\bibfnamefont {Y.}~\bibnamefont
  {Tokura}}\ and\ \bibinfo {author} {\bibfnamefont {S.}~\bibnamefont {Seki}},\
  }\bibfield  {title} {\bibinfo {title} {\textit{Multiferroics with spiral spin
  orders}},\ }\href {https://doi.org/10.1002/adma.200901961} {\bibfield
  {journal} {\bibinfo  {journal} {Adv. Mater.}\ }\textbf {\bibinfo {volume}
  {22}},\ \bibinfo {pages} {1554} (\bibinfo {year} {2010})}\BibitemShut
  {NoStop}%
\bibitem [{\citenamefont {Gu}\ \emph {et~al.}(2018)\citenamefont {Gu},
  \citenamefont {Zhao}, \citenamefont {Chen}, \citenamefont {Lee},
  \citenamefont {Choi}, \citenamefont {Han}, \citenamefont {Ling},
  \citenamefont {Pi}, \citenamefont {Zhang}, \citenamefont {Chen},
  \citenamefont {Yang}, \citenamefont {Zhou},\ and\ \citenamefont
  {Sun}}]{GuZhao18}%
  \BibitemOpen
  \bibfield  {author} {\bibinfo {author} {\bibfnamefont {C.~C.}\ \bibnamefont
  {Gu}}, \bibinfo {author} {\bibfnamefont {Z.~Y.}\ \bibnamefont {Zhao}},
  \bibinfo {author} {\bibfnamefont {X.~L.}\ \bibnamefont {Chen}}, \bibinfo
  {author} {\bibfnamefont {M.}~\bibnamefont {Lee}}, \bibinfo {author}
  {\bibfnamefont {E.~S.}\ \bibnamefont {Choi}}, \bibinfo {author}
  {\bibfnamefont {Y.~Y.}\ \bibnamefont {Han}}, \bibinfo {author} {\bibfnamefont
  {L.~S.}\ \bibnamefont {Ling}}, \bibinfo {author} {\bibfnamefont
  {L.}~\bibnamefont {Pi}}, \bibinfo {author} {\bibfnamefont {Y.~H.}\
  \bibnamefont {Zhang}}, \bibinfo {author} {\bibfnamefont {G.}~\bibnamefont
  {Chen}}, \bibinfo {author} {\bibfnamefont {Z.~R.}\ \bibnamefont {Yang}},
  \bibinfo {author} {\bibfnamefont {H.~D.}\ \bibnamefont {Zhou}},\ and\
  \bibinfo {author} {\bibfnamefont {X.~F.}\ \bibnamefont {Sun}},\ }\bibfield
  {title} {\bibinfo {title} {\textit{Field-driven quantum criticality in the
  spinel magnet ZnCr$_\text{\!2}$Se$_\text{4}$}},\ }\href
  {https://doi.org/10.1103/PhysRevLett.120.147204} {\bibfield  {journal}
  {\bibinfo  {journal} {Phys. Rev. Lett.}\ }\textbf {\bibinfo {volume} {120}},\
  \bibinfo {pages} {147204} (\bibinfo {year} {2018})}\BibitemShut {NoStop}%
\bibitem [{\citenamefont {Inosov}\ \emph {et~al.}(2018)\citenamefont {Inosov}
  \emph {et~al.}}]{DataLET}%
  \BibitemOpen
  \bibfield  {author} {\bibinfo {author} {\bibfnamefont {D.~S.}\ \bibnamefont
  {Inosov}} \emph {et~al.},\ }\href@noop {} {\bibinfo {title} {\textit{Nature
  of the new field-induced magnetic phase in
  ZnCr$_\text{\!2}$Se$_\text{4}$}}},\ \bibinfo {howpublished} {STFC ISIS
  Neutron and Muon Source,
  \href{https://doi.org/10.5286/ISIS.E.RB1810109}{doi:10.5286/ISIS.E.RB1810109}}
  (\bibinfo {year} {2018})\BibitemShut {NoStop}%
\bibitem [{\citenamefont {Bewley}\ \emph {et~al.}(2011)\citenamefont {Bewley},
  \citenamefont {Taylor},\ and\ \citenamefont {Bennington}}]{BewleyTaylor11}%
  \BibitemOpen
  \bibfield  {author} {\bibinfo {author} {\bibfnamefont {R.~I.}\ \bibnamefont
  {Bewley}}, \bibinfo {author} {\bibfnamefont {J.~W.}\ \bibnamefont {Taylor}},\
  and\ \bibinfo {author} {\bibfnamefont {S.~M.}\ \bibnamefont {Bennington}},\
  }\bibfield  {title} {\bibinfo {title} {\textit{LET, a cold-neutron multi-disk
  chopper spectrometer at ISIS}},\ }\href
  {https://doi.org/10.1016/j.nima.2011.01.173} {\bibfield  {journal} {\bibinfo
  {journal} {Nucl. Instrum. Methods Phys. Res. Sect. A\,--\,Accel. Spectrom.
  Dect. Assoc. Equip.}\ }\textbf {\bibinfo {volume} {637}},\ \bibinfo {pages}
  {128} (\bibinfo {year} {2011})}\BibitemShut {NoStop}%
\bibitem [{\citenamefont {Boehm}\ \emph {et~al.}(2008)\citenamefont {Boehm},
  \citenamefont {Hiess}, \citenamefont {Kulda}, \citenamefont {Roux},\ and\
  \citenamefont {Saroun}}]{BoehmHiess08}%
  \BibitemOpen
  \bibfield  {author} {\bibinfo {author} {\bibfnamefont {M.}~\bibnamefont
  {Boehm}}, \bibinfo {author} {\bibfnamefont {A.}~\bibnamefont {Hiess}},
  \bibinfo {author} {\bibfnamefont {J.}~\bibnamefont {Kulda}}, \bibinfo
  {author} {\bibfnamefont {S.}~\bibnamefont {Roux}},\ and\ \bibinfo {author}
  {\bibfnamefont {J.}~\bibnamefont {Saroun}},\ }\bibfield  {title} {\bibinfo
  {title} {\textit{ThALES\,---\,three-axis low-energy spectroscopy at the
  Institut Laue-Langevin}},\ }\href
  {https://doi.org/10.1088/0957-0233/19/3/034024} {\bibfield  {journal}
  {\bibinfo  {journal} {Meas. Sci. Technol.}\ }\textbf {\bibinfo {volume}
  {19}},\ \bibinfo {pages} {034024} (\bibinfo {year} {2008})}\BibitemShut
  {NoStop}%
\bibitem [{\citenamefont {Boehm}\ \emph {et~al.}(2015)\citenamefont {Boehm},
  \citenamefont {Steffens}, \citenamefont {Kulda}, \citenamefont {Klicpera},
  \citenamefont {Roux}, \citenamefont {Courtois}, \citenamefont {Svoboda},
  \citenamefont {Saroun},\ and\ \citenamefont {Sechovsky}}]{BoehmSteffens15}%
  \BibitemOpen
  \bibfield  {author} {\bibinfo {author} {\bibfnamefont {M.}~\bibnamefont
  {Boehm}}, \bibinfo {author} {\bibfnamefont {P.}~\bibnamefont {Steffens}},
  \bibinfo {author} {\bibfnamefont {J.}~\bibnamefont {Kulda}}, \bibinfo
  {author} {\bibfnamefont {M.}~\bibnamefont {Klicpera}}, \bibinfo {author}
  {\bibfnamefont {S.}~\bibnamefont {Roux}}, \bibinfo {author} {\bibfnamefont
  {P.}~\bibnamefont {Courtois}}, \bibinfo {author} {\bibfnamefont
  {P.}~\bibnamefont {Svoboda}}, \bibinfo {author} {\bibfnamefont
  {J.}~\bibnamefont {Saroun}},\ and\ \bibinfo {author} {\bibfnamefont
  {V.}~\bibnamefont {Sechovsky}},\ }\bibfield  {title} {\bibinfo {title}
  {\textit{ThALES\,---\,three-axis low-energy spectroscopy for highly
  correlated electron systems}},\ }\href
  {https://doi.org/10.1080/10448632.2015.1057050} {\bibfield  {journal}
  {\bibinfo  {journal} {Neutron News}\ }\textbf {\bibinfo {volume} {26}},\
  \bibinfo {pages} {18} (\bibinfo {year} {2015})}\BibitemShut {NoStop}%
\bibitem [{\citenamefont {Tymoshenko}\ \emph {et~al.}()\citenamefont
  {Tymoshenko} \emph {et~al.}}]{DataThales}%
  \BibitemOpen
  \bibfield  {author} {\bibinfo {author} {\bibfnamefont {Y.~V.}\ \bibnamefont
  {Tymoshenko}} \emph {et~al.},\ }\href@noop {} {\bibinfo {title}
  {\textit{Non-monotonic behavior of the spin gap in
  ZnCr$_\text{\!2}$Se$_\text{4}$ in magnetic field}}},\ \bibinfo {howpublished}
  {Institut Laue-Langevin (ILL),
  \href{https://doi.ill.fr/10.5291/ILL-DATA.4-05-734}{doi:10.5291/ILL-DATA.4-05-734}}\BibitemShut
  {NoStop}%
\bibitem [{\citenamefont {Rotter}\ \emph {et~al.}(1998)\citenamefont {Rotter},
  \citenamefont {M\"uller}, \citenamefont {Gratz}, \citenamefont {Doerr},\ and\
  \citenamefont {Loewenhaupt}}]{RotterMueller98}%
  \BibitemOpen
  \bibfield  {author} {\bibinfo {author} {\bibfnamefont {M.}~\bibnamefont
  {Rotter}}, \bibinfo {author} {\bibfnamefont {H.}~\bibnamefont {M\"uller}},
  \bibinfo {author} {\bibfnamefont {E.}~\bibnamefont {Gratz}}, \bibinfo
  {author} {\bibfnamefont {M.}~\bibnamefont {Doerr}},\ and\ \bibinfo {author}
  {\bibfnamefont {M.}~\bibnamefont {Loewenhaupt}},\ }\bibfield  {title}
  {\bibinfo {title} {\textit{A miniature capacitance dilatometer for thermal
  expansion and magnetostriction}},\ }\href {https://doi.org/10.1063/1.1149009}
  {\bibfield  {journal} {\bibinfo  {journal} {Rev. Sci. Instrum.}\ }\textbf
  {\bibinfo {volume} {69}},\ \bibinfo {pages} {2742} (\bibinfo {year}
  {1998})}\BibitemShut {NoStop}%
\bibitem [{\citenamefont {Park}\ \emph {et~al.}(2019)\citenamefont {Park},
  \citenamefont {Kwon}, \citenamefont {Lee}, \citenamefont {Khim},
  \citenamefont {Bhoi}, \citenamefont {Park},\ and\ \citenamefont
  {Kim}}]{ParkKwon19}%
  \BibitemOpen
  \bibfield  {author} {\bibinfo {author} {\bibfnamefont {S.}~\bibnamefont
  {Park}}, \bibinfo {author} {\bibfnamefont {S.}~\bibnamefont {Kwon}}, \bibinfo
  {author} {\bibfnamefont {S.}~\bibnamefont {Lee}}, \bibinfo {author}
  {\bibfnamefont {S.}~\bibnamefont {Khim}}, \bibinfo {author} {\bibfnamefont
  {D.}~\bibnamefont {Bhoi}}, \bibinfo {author} {\bibfnamefont {C.~B.}\
  \bibnamefont {Park}},\ and\ \bibinfo {author} {\bibfnamefont {K.~H.}\
  \bibnamefont {Kim}},\ }\bibfield  {title} {\bibinfo {title}
  {\textit{Interactions in the bond-frustrated helimagnet
  {Z}n{C}r$_{\!2}${S}e$_{4}$ investigated by NMR}},\ }\href
  {https://doi.org/10.1038/s41598-019-52962-4} {\bibfield  {journal} {\bibinfo
  {journal} {Sci. Rep.}\ }\textbf {\bibinfo {volume} {9}},\ \bibinfo {pages}
  {16627} (\bibinfo {year} {2019})}\BibitemShut {NoStop}%
\bibitem [{\citenamefont {Takahashi}\ and\ \citenamefont
  {Shimizu}(1977)}]{TakahashiShimizu77}%
  \BibitemOpen
  \bibfield  {author} {\bibinfo {author} {\bibfnamefont {Y.}~\bibnamefont
  {Takahashi}}\ and\ \bibinfo {author} {\bibfnamefont {M.}~\bibnamefont
  {Shimizu}},\ }\bibfield  {title} {\bibinfo {title} {\textit{Anisotropy of
  paramagnetostriction of metals with HCP structure}},\ }\href
  {https://doi.org/10.1088/0305-4608/7/3/017} {\bibfield  {journal} {\bibinfo
  {journal} {J.~Phys.~F: Metal Phys.}\ }\textbf {\bibinfo {volume} {7}},\
  \bibinfo {pages} {471} (\bibinfo {year} {1977})}\BibitemShut {NoStop}%
\bibitem [{\citenamefont {{Etienne du Tremolet de
  Lacheisserie}}(1993)}]{TremoletdeLacheisserie93}%
  \BibitemOpen
  \bibfield  {author} {\bibinfo {author} {\bibnamefont {{Etienne du Tremolet de
  Lacheisserie}}},\ }\href@noop {} {\emph {\bibinfo {title} {Magnetostriction:
  Theory and Applications of Magnetoelasticity}}}\ (\bibinfo  {publisher} {CRC
  Press},\ \bibinfo {address} {Boca Raton, FL},\ \bibinfo {year}
  {1993})\BibitemShut {NoStop}%
\bibitem [{\citenamefont {Cullity}\ and\ \citenamefont
  {Graham}(2009)}]{CullityGraham09}%
  \BibitemOpen
  \bibfield  {author} {\bibinfo {author} {\bibfnamefont {B.~D.}\ \bibnamefont
  {Cullity}}\ and\ \bibinfo {author} {\bibfnamefont {C.~D.}\ \bibnamefont
  {Graham}},\ }\href@noop {} {\emph {\bibinfo {title} {Introduction to Magnetic
  Materials}}},\ \bibinfo {edition} {2nd}\ ed.,\ edited by\ \bibinfo {editor}
  {\bibfnamefont {L.}~\bibnamefont {Hanzo}}\ (\bibinfo  {publisher} {IEEE
  Press, Wiley},\ \bibinfo {address} {Hoboken, NJ},\ \bibinfo {year} {2009})\
  \bibinfo {note} {\hspace{-2.8pt}, chap.~8}\BibitemShut {NoStop}%
\bibitem [{\citenamefont {Croft}\ \emph {et~al.}(1978)\citenamefont {Croft},
  \citenamefont {Zoric},\ and\ \citenamefont {Parks}}]{CroftZoric78}%
  \BibitemOpen
  \bibfield  {author} {\bibinfo {author} {\bibfnamefont {M.}~\bibnamefont
  {Croft}}, \bibinfo {author} {\bibfnamefont {I.}~\bibnamefont {Zoric}},\ and\
  \bibinfo {author} {\bibfnamefont {R.~D.}\ \bibnamefont {Parks}},\ }\bibfield
  {title} {\bibinfo {title} {\textit{Anisotropic magnetostriction of
  CeAl$_\text{2}$ near its antiferromagnetic transition}},\ }\href
  {https://doi.org/10.1103/PhysRevB.18.345} {\bibfield  {journal} {\bibinfo
  {journal} {Phys. Rev.~B}\ }\textbf {\bibinfo {volume} {18}},\ \bibinfo
  {pages} {345} (\bibinfo {year} {1978})}\BibitemShut {NoStop}%
\bibitem [{\citenamefont {Zieglowski}\ \emph {et~al.}(1987)\citenamefont
  {Zieglowski}, \citenamefont {Wohlleben}, \citenamefont {Schmidt},
  \citenamefont {M\"uller-Hartmann},\ and\ \citenamefont
  {Winzer}}]{ZieglowskiWohlleben87}%
  \BibitemOpen
  \bibfield  {author} {\bibinfo {author} {\bibfnamefont {J.}~\bibnamefont
  {Zieglowski}}, \bibinfo {author} {\bibfnamefont {D.}~\bibnamefont
  {Wohlleben}}, \bibinfo {author} {\bibfnamefont {H.~J.}\ \bibnamefont
  {Schmidt}}, \bibinfo {author} {\bibfnamefont {E.}~\bibnamefont
  {M\"uller-Hartmann}},\ and\ \bibinfo {author} {\bibfnamefont
  {K.}~\bibnamefont {Winzer}},\ }\bibfield  {title} {\bibinfo {title}
  {\textit{Magnetostriction of Ce in Ce$_x$La$_{\text{1}-x}$Al$_\text{2}$}},\
  }\href {https://doi.org/10.1103/PhysRevB.35.8595} {\bibfield  {journal}
  {\bibinfo  {journal} {Phys. Rev. B}\ }\textbf {\bibinfo {volume} {35}},\
  \bibinfo {pages} {8595} (\bibinfo {year} {1987})}\BibitemShut {NoStop}%
\bibitem [{\citenamefont {Fawcett}\ \emph {et~al.}(1991)\citenamefont
  {Fawcett}, \citenamefont {Pluzhnikov},\ and\ \citenamefont
  {Klimker}}]{FawcettPluzhnikov91}%
  \BibitemOpen
  \bibfield  {author} {\bibinfo {author} {\bibfnamefont {E.}~\bibnamefont
  {Fawcett}}, \bibinfo {author} {\bibfnamefont {V.}~\bibnamefont
  {Pluzhnikov}},\ and\ \bibinfo {author} {\bibfnamefont {H.}~\bibnamefont
  {Klimker}},\ }\bibfield  {title} {\bibinfo {title} {\textit{Thermal expansion
  and magnetostriction of CeAl$_\text{2}$}},\ }\href
  {https://doi.org/10.1103/PhysRevB.43.8531} {\bibfield  {journal} {\bibinfo
  {journal} {Phys. Rev.~B}\ }\textbf {\bibinfo {volume} {43}},\ \bibinfo
  {pages} {8531} (\bibinfo {year} {1991})}\BibitemShut {NoStop}%
\bibitem [{\citenamefont {Sokolov}\ \emph {et~al.}(1999)\citenamefont
  {Sokolov}, \citenamefont {Nakamura},\ and\ \citenamefont
  {Shiga}}]{SokolovNakamura99}%
  \BibitemOpen
  \bibfield  {author} {\bibinfo {author} {\bibfnamefont {A.~Y.}\ \bibnamefont
  {Sokolov}}, \bibinfo {author} {\bibfnamefont {H.}~\bibnamefont {Nakamura}},\
  and\ \bibinfo {author} {\bibfnamefont {M.}~\bibnamefont {Shiga}},\ }\bibfield
   {title} {\bibinfo {title} {\textit{Magnetostriction of ytterbium-based Kondo
  compounds Yb$X$Cu$_\text{4}$ ($X$~=~In, Ag and Au)}},\ }\href
  {https://doi.org/10.1088/0953-8984/11/33/314} {\bibfield  {journal} {\bibinfo
   {journal} {J.~Phys: Condens. Matter}\ }\textbf {\bibinfo {volume} {11}},\
  \bibinfo {pages} {6463} (\bibinfo {year} {1999})}\BibitemShut {NoStop}%
\bibitem [{\citenamefont {Gu}\ \emph {et~al.}(2016)\citenamefont {Gu},
  \citenamefont {Yang}, \citenamefont {Chen}, \citenamefont {Pi},\ and\
  \citenamefont {Zhang}}]{GuYang16}%
  \BibitemOpen
  \bibfield  {author} {\bibinfo {author} {\bibfnamefont {C.}~\bibnamefont
  {Gu}}, \bibinfo {author} {\bibfnamefont {Z.}~\bibnamefont {Yang}}, \bibinfo
  {author} {\bibfnamefont {X.}~\bibnamefont {Chen}}, \bibinfo {author}
  {\bibfnamefont {L.}~\bibnamefont {Pi}},\ and\ \bibinfo {author}
  {\bibfnamefont {Y.}~\bibnamefont {Zhang}},\ }\bibfield  {title} {\bibinfo
  {title} {\textit{Negative thermal expansion and magnetostriction in the
  frustrated spinel ZnCr$_\text{2}$(Se$_{\text{1}-x}$S$_x$)}},\ }\href
  {https://doi.org/10.1088/0953-8984/28/18/18LT01} {\bibfield  {journal}
  {\bibinfo  {journal} {J.~Phys.: Condens. Matter}\ }\textbf {\bibinfo {volume}
  {28}},\ \bibinfo {pages} {18LT01} (\bibinfo {year} {2016})}\BibitemShut
  {NoStop}%
\bibitem [{\citenamefont {Rau}\ \emph {et~al.}(2018)\citenamefont {Rau},
  \citenamefont {McClarty},\ and\ \citenamefont {Moessner}}]{RauMcClarty18}%
  \BibitemOpen
  \bibfield  {author} {\bibinfo {author} {\bibfnamefont {J.~G.}\ \bibnamefont
  {Rau}}, \bibinfo {author} {\bibfnamefont {P.~A.}\ \bibnamefont {McClarty}},\
  and\ \bibinfo {author} {\bibfnamefont {R.}~\bibnamefont {Moessner}},\
  }\bibfield  {title} {\bibinfo {title} {\textit{Pseudo-Goldstone gaps and
  order-by-quantum disorder in frustrated magnets}},\ }\href
  {https://doi.org/10.1103/PhysRevLett.121.237201} {\bibfield  {journal}
  {\bibinfo  {journal} {Phys. Rev. Lett.}\ }\textbf {\bibinfo {volume} {121}},\
  \bibinfo {pages} {237201} (\bibinfo {year} {2018})}\BibitemShut {NoStop}%
\bibitem [{\citenamefont {Villain}\ \emph {et~al.}(1980)\citenamefont
  {Villain}, \citenamefont {Bidaux}, \citenamefont {Carton},\ and\
  \citenamefont {Conte}}]{VillainBidaux80}%
  \BibitemOpen
  \bibfield  {author} {\bibinfo {author} {\bibfnamefont {J.}~\bibnamefont
  {Villain}}, \bibinfo {author} {\bibfnamefont {R.}~\bibnamefont {Bidaux}},
  \bibinfo {author} {\bibfnamefont {J.-P.}\ \bibnamefont {Carton}},\ and\
  \bibinfo {author} {\bibfnamefont {R.}~\bibnamefont {Conte}},\ }\bibfield
  {title} {\bibinfo {title} {\textit{Order as an effect of disorder}},\ }\href
  {https://doi.org/10.1051/jphys:0198000410110126300} {\bibfield  {journal}
  {\bibinfo  {journal} {J~Phys. France}\ }\textbf {\bibinfo {volume} {41}},\
  \bibinfo {pages} {1263} (\bibinfo {year} {1980})}\BibitemShut {NoStop}%
\bibitem [{\citenamefont {Henley}(1989)}]{Henley89}%
  \BibitemOpen
  \bibfield  {author} {\bibinfo {author} {\bibfnamefont {C.~L.}\ \bibnamefont
  {Henley}},\ }\bibfield  {title} {\bibinfo {title} {Ordering due to disorder
  in a frustrated vector antiferromagnet},\ }\href
  {https://doi.org/10.1103/PhysRevLett.62.2056} {\bibfield  {journal} {\bibinfo
   {journal} {Phys. Rev. Lett.}\ }\textbf {\bibinfo {volume} {62}},\ \bibinfo
  {pages} {2056} (\bibinfo {year} {1989})}\BibitemShut {NoStop}%
\bibitem [{\citenamefont {Sun}()}]{XFSunPrivate}%
  \BibitemOpen
  \bibfield  {author} {\bibinfo {author} {\bibfnamefont {X.}~\bibnamefont
  {Sun}},\ }\href@noop {} {}\bibinfo {note} {{private
  communication}}\BibitemShut {NoStop}%
\bibitem [{\citenamefont {Portnichenko}\ \emph {et~al.}(2016)\citenamefont
  {Portnichenko}, \citenamefont {Romh\'anyi}, \citenamefont {Onykiienko},
  \citenamefont {Henschel}, \citenamefont {Schmidt}, \citenamefont {Cameron},
  \citenamefont {Surmach}, \citenamefont {Lim}, \citenamefont {Park},
  \citenamefont {Schneidewind}, \citenamefont {Abernathy}, \citenamefont
  {Rosner}, \citenamefont {van~den Brink},\ and\ \citenamefont
  {Inosov}}]{PortnichenkoRomhanyi16}%
  \BibitemOpen
  \bibfield  {author} {\bibinfo {author} {\bibfnamefont {P.~Y.}\ \bibnamefont
  {Portnichenko}}, \bibinfo {author} {\bibfnamefont {J.}~\bibnamefont
  {Romh\'anyi}}, \bibinfo {author} {\bibfnamefont {Y.~A.}\ \bibnamefont
  {Onykiienko}}, \bibinfo {author} {\bibfnamefont {A.}~\bibnamefont
  {Henschel}}, \bibinfo {author} {\bibfnamefont {M.}~\bibnamefont {Schmidt}},
  \bibinfo {author} {\bibfnamefont {A.~S.}\ \bibnamefont {Cameron}}, \bibinfo
  {author} {\bibfnamefont {M.~A.}\ \bibnamefont {Surmach}}, \bibinfo {author}
  {\bibfnamefont {J.~A.}\ \bibnamefont {Lim}}, \bibinfo {author} {\bibfnamefont
  {J.~T.}\ \bibnamefont {Park}}, \bibinfo {author} {\bibfnamefont
  {A.}~\bibnamefont {Schneidewind}}, \bibinfo {author} {\bibfnamefont {D.~L.}\
  \bibnamefont {Abernathy}}, \bibinfo {author} {\bibfnamefont {H.}~\bibnamefont
  {Rosner}}, \bibinfo {author} {\bibfnamefont {J.}~\bibnamefont {van~den
  Brink}},\ and\ \bibinfo {author} {\bibfnamefont {D.~S.}\ \bibnamefont
  {Inosov}},\ }\bibfield  {title} {\bibinfo {title} {\textit{Magnon spectrum of
  the helimagnetic insulator Cu$_\text{2}$OSeO$_\text{3}$}},\ }\href
  {https://doi.org/10.1038/ncomms10725} {\bibfield  {journal} {\bibinfo
  {journal} {Nat. Commun.}\ }\textbf {\bibinfo {volume} {7}},\ \bibinfo {pages}
  {10725} (\bibinfo {year} {2016})}\BibitemShut {NoStop}%
\bibitem [{\citenamefont {Prasai}\ \emph {et~al.}(2017)\citenamefont {Prasai},
  \citenamefont {Trump}, \citenamefont {Marcus}, \citenamefont {Akopyan},
  \citenamefont {Huang}, \citenamefont {McQueen},\ and\ \citenamefont
  {Cohn}}]{PrasaiTrump17}%
  \BibitemOpen
  \bibfield  {author} {\bibinfo {author} {\bibfnamefont {N.}~\bibnamefont
  {Prasai}}, \bibinfo {author} {\bibfnamefont {B.~A.}\ \bibnamefont {Trump}},
  \bibinfo {author} {\bibfnamefont {G.~G.}\ \bibnamefont {Marcus}}, \bibinfo
  {author} {\bibfnamefont {A.}~\bibnamefont {Akopyan}}, \bibinfo {author}
  {\bibfnamefont {S.~X.}\ \bibnamefont {Huang}}, \bibinfo {author}
  {\bibfnamefont {T.~M.}\ \bibnamefont {McQueen}},\ and\ \bibinfo {author}
  {\bibfnamefont {J.~L.}\ \bibnamefont {Cohn}},\ }\bibfield  {title} {\bibinfo
  {title} {\textit{Ballistic magnon heat conduction and possible Poiseuille
  flow in the helimagnetic insulator Cu$_\text{2}$OSeO$_\text{3}$}},\ }\href
  {https://doi.org/10.1103/PhysRevB.95.224407} {\bibfield  {journal} {\bibinfo
  {journal} {Phys. Rev.~B}\ }\textbf {\bibinfo {volume} {95}},\ \bibinfo
  {pages} {224407} (\bibinfo {year} {2017})}\BibitemShut {NoStop}%
\bibitem [{\citenamefont {Streib}\ \emph {et~al.}(2019)\citenamefont {Streib},
  \citenamefont {Vidal-Silva}, \citenamefont {Shen},\ and\ \citenamefont
  {Bauer}}]{StreibVidalSilva19}%
  \BibitemOpen
  \bibfield  {author} {\bibinfo {author} {\bibfnamefont {S.}~\bibnamefont
  {Streib}}, \bibinfo {author} {\bibfnamefont {N.}~\bibnamefont {Vidal-Silva}},
  \bibinfo {author} {\bibfnamefont {K.}~\bibnamefont {Shen}},\ and\ \bibinfo
  {author} {\bibfnamefont {G.~E.~W.}\ \bibnamefont {Bauer}},\ }\bibfield
  {title} {\bibinfo {title} {\textit{Magnon-phonon interactions in magnetic
  insulators}},\ }\href {https://doi.org/10.1103/PhysRevB.99.184442} {\bibfield
   {journal} {\bibinfo  {journal} {Phys. Rev. B}\ }\textbf {\bibinfo {volume}
  {99}},\ \bibinfo {pages} {184442} (\bibinfo {year} {2019})}\BibitemShut
  {NoStop}%
\bibitem [{SM()}]{SM}%
  \BibitemOpen
  \href@noop {} {}\bibinfo {note} {{See Supplemental Material at [URL will be
  inserted by publisher] for further details about the calculation of the
  phonon-magnon scattering rates, lattice and magnon thermal
  conductivities.}}\BibitemShut {Stop}%
\end{thebibliography}%

\onecolumngrid\clearpage

\pagestyle{plain}

\renewcommand\thefigure{S\arabic{figure}}
\renewcommand\thetable{S\arabic{table}}
\renewcommand\theequation{S\arabic{equation}}
\renewcommand\thesection{S\arabic{section}}
\renewcommand\thesubsection{S\arabic{subsection}}
\renewcommand\bibsection{\section*{\sffamily\bfseries\footnotesize Supplementary References\vspace{-6pt}\hfill~}}

\citestyle{supplement}

\setcounter{page}{1}\setcounter{figure}{0}\setcounter{table}{0}\setcounter{equation}{0}\setcounter{section}{0}

\pagestyle{plain}
\makeatletter
\renewcommand{\@oddfoot}{\hfill\scriptsize\textsf{--~S\thepage~--}\hfill}
\renewcommand{\@evenfoot}{\hfill\scriptsize\textsf{--~S\thepage~--}\hfill}
\makeatother


\normalsize

\begin{center}
\Large{Supplemental Material for:}\bigskip\\
\large{\sl\textbf{``Magnetic-field dependence of low-energy magnons, anisotropic heat conduction, and spontaneous relaxation of magnetic domains in the cubic helimagnet ZnCr$_\text{2}$Se$_\text{4}$''}}\bigskip\\
\normalsize{D.~S. Inosov, Y.~O. Onykiienko, Y.~V. Tymoshenko, A.~Akopyan, D.~P. Shukla, N.~Prasai,\\ M.~Doerr, S.~Zherlitsyn, D.~Voneshen, M.~Boehm, V.~Tsurkan, V.~Felea, A.~Loidl, and J.~L. Cohn}
\end{center}

{\twocolumngrid

\subsection{Magnon dispersion}\vspace{-2pt}

In prior work~\cite{TymoshenkoOnykiienko17} an analytical form for the $B=0$ magnon dispersion was developed with parameters constrained by fitting to the inelastic neutron scattering results along the main symmetry directions. Figure~\ref{Fig:Dispersions} shows the results from this model in the $k_x-k_y$ and $k_x-k_z$ planes, extended to applied fields $B = 3$, 6, 8~T based on the scattering data presented in Sec.~\ref{Sec:INS}. The axes are chosen with $k_z$ parallel to the spin spiral axis (designated $\mathbf{q}_\text{h}$ in the main text); the polar and azimuthal angles, $\theta$ and $\phi$, are labeled in the $B=0$ diagrams.

\begin{figure}[b]
\centerline{\includegraphics[width=\linewidth]{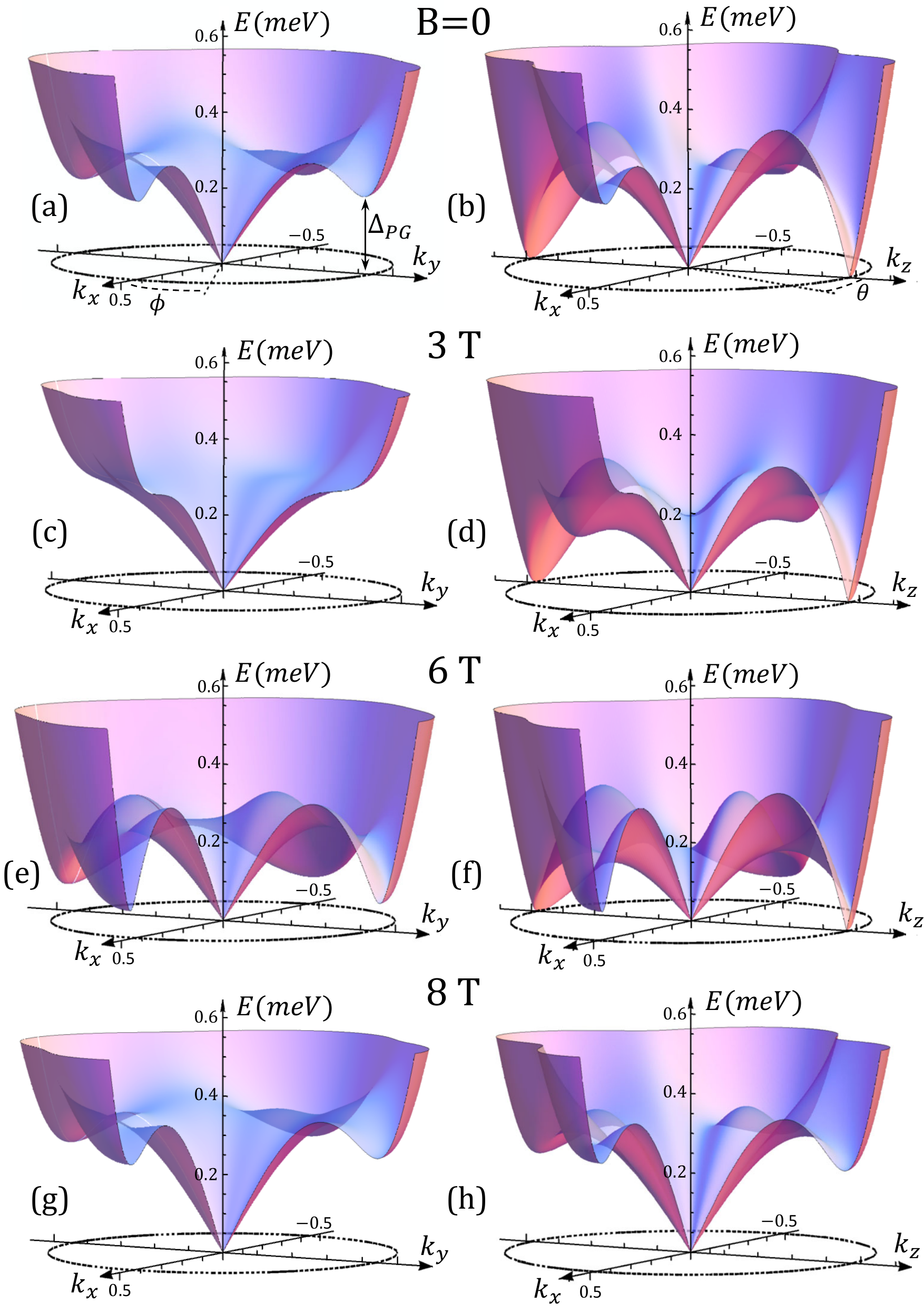}}
\caption{Three-dimensional magnon dispersions, shown in the $k_x$-$k_y$ (left) and $k_x$-$k_z$ (right) planes for four values of magnetic field. Wave numbers are in units of $2\pi/a$. The polar and azimuthal angles, $\theta, \phi$, and the pseudo-Goldstone magnon gap ($\Delta_{\rm PG}$, see Fig.~\ref{fig:TAS}) are labeled in the plots for $B=0$.}
\label{Fig:Dispersions}
\end{figure}

These approximations to the three-dimensional dispersions $E(\phi,\theta,k)$ were employed in calculations of the phonon-magnon scattering rate and magnon thermal conductivity as described further below.

\subsection{Phonon-magnon scattering rate}\vspace{-2pt}

Following the recent work of Streib \textit{et al.}~\cite{StreibVidalSilva19}, we derived the scattering rate for 1-phonon, 2-magnon scattering as:
$$
\frac{1}{\tau_{\mathbf{q},\lambda}}=\frac{\pi\hbar}{mN}\sum_{\mathbf{k}}\frac{|\Gamma_{\mathbf{k,\,q-k},\,\lambda}|^{2\hspace{-2pt}}}{\varepsilon_{\mathbf{q},\lambda}}\left(1{\kern-1pt}+{\kern-1pt}n_{\mathbf{q-k}}{\kern-1pt}+{\kern-1pt}n_{\mathbf{k}}\right)
\delta\!\left(E_{\mathbf{k}}\!+\!E_{\mathbf{q-k}}\!+\!{\varepsilon_{\mathbf{q},\lambda}}\right)\!,
$$
where $m$ and $N$ are the mass per unit cell and the number of unit cells, respectively,  $\mathbf{k}$ and $\mathbf{q}$ are the magnon and phonon wave vectors ($\lambda$ is a polarization index), $n_{\mathbf{k}}$ is the Bose-Einstein distribution function, $E_{\mathbf{k}}$ and $\varepsilon_{\mathbf{q},\lambda}=\hbar c_{\lambda}|\mathbf{q}|$ are magnon and phonon energies, respectively, and the $\delta$-function enforces energy conservation. We used an approximate analytic expression for $E_{\mathbf{k}}$ with parameters constrained by neutron scattering measurements along the main symmetry directions (see below). $\Gamma_{\mathbf{k,\,q-k},\,\lambda}$ is the matrix element for phonon-magnon interaction that is linear in $q$ with a scale set by the magnetoelastic constants. The latter were assumed to be isotropic, and only transverse phonons were considered (they should predominate in the scattering with magnons given their lower energies), with both modes taken to have the same velocity, $c_{\lambda}\equiv c=2$~km/s.

\vspace{-2pt}\subsection{Calculation of the phonon-magnon scattering rate}\vspace{-2pt}

The sums over magnon momenta $\mathbf{k}$ were performed over a uniform $k$-space mesh ($120\times 120 \times 120$) with spacing 0.01 in the range from $-0.595$ to 0.595 [in reciprocal lattice units, $q/(2\pi/a$)]. The scattering rates were found to be independent of $\phi$ and very weakly dependent on $\theta$. $\tau_{\mathbf{q}}^{-1}$ was computed for phonon coordinates $q$ restricted to relative values in the range 0 to 0.422 with spacing 0.002 for the $q$ range 0--0.104 and 0.008 for the remaining range. The subsequent integrations to determine $\kappa$ employed interpolations on $\tau_{\mathbf{q}}^{-1}$. The energy $\delta$ function was approximated~\cite{StreibVidalSilva19} as $\delta(\epsilon)\approx (1/\sqrt{\pi}\alpha)\exp(-\epsilon^2/\alpha^2)$, with $\alpha=0.005$~meV.

The absence of anisotropy ($\theta$ dependence) in the scattering rate is a consequence of the peculiar magnon dispersion. The dominant magnons in the scattering of phonons at low $T$ are those with momenta $k \approx 0.3$ and energies in the range $E\approx 0.15$--0.3~meV restricted to narrow conical regions oriented along the coordinate axes where the dispersion is flat and the magnon density of states correspondingly high (Fig.~\ref{Fig:Dispersions}). Energy conservation dictates $q\ll k$ and thus momentum conservation, $\mathbf{k}+\mathbf{k^{\prime}}=\mathbf{q}$, is achieved with magnon states having nearly opposing momenta, either along or transverse to $\mathbf{q}$.

\begin{figure}[t]
\centerline{\includegraphics[width=\linewidth]{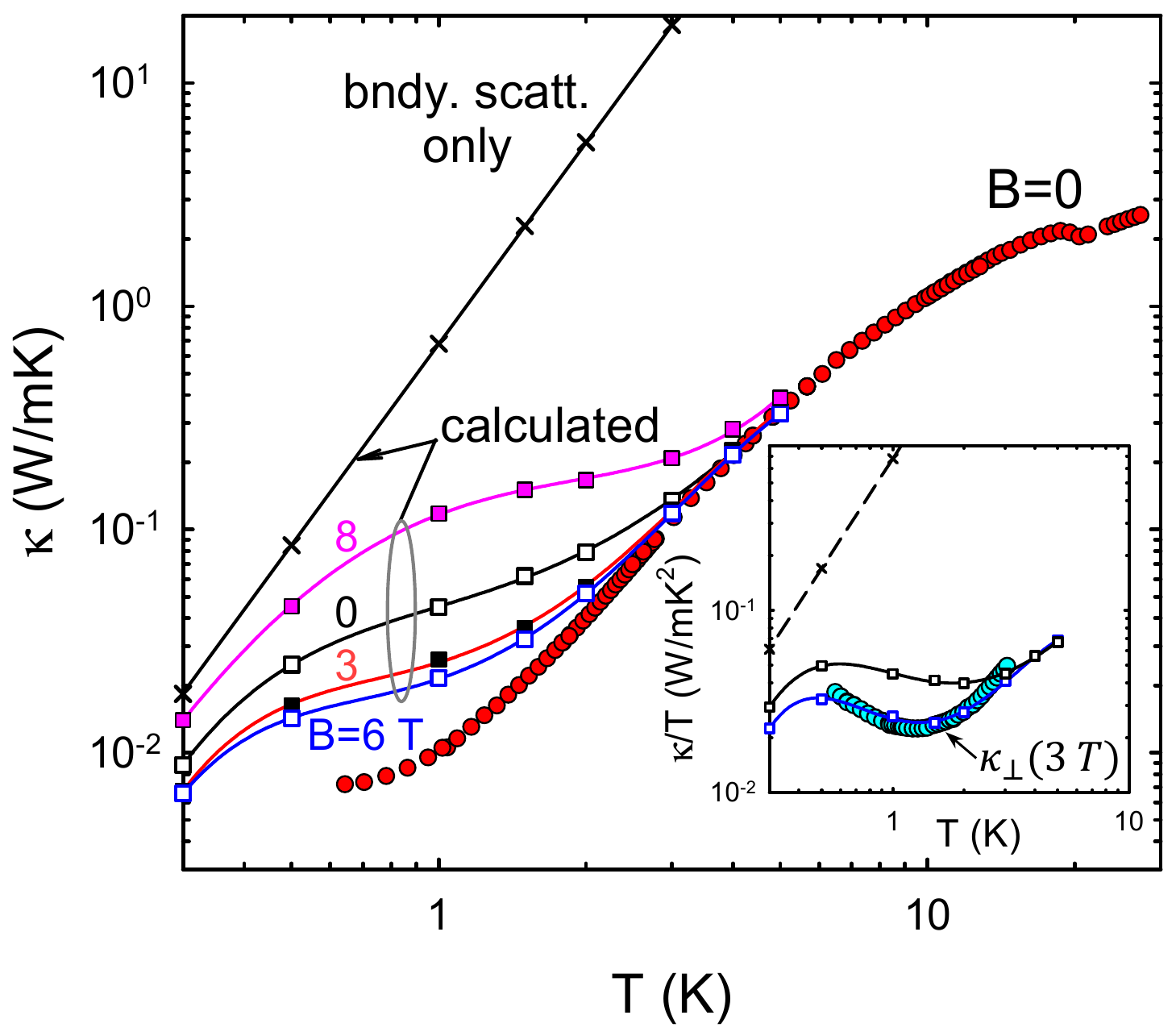}}
\caption{Computed thermal lattice conductivities (open squares; $\kappa_{\rm L,\bot}=\kappa_{\rm L,\|}$ to 1\%) at various fields compared to the experimental $B=0$ data (filled circles). The magnetoelastic constant was set so that the $T=5$~K conductivities matched the experimental data. Also shown is the computed thermal conductivity with only the boundary scattering term included ($\times$). Inset: Computed $\kappa_{\rm L}/T$ for the boundary scattering and $B=0$, 3~T curves from the main panel and corresponding experimental 3~T data (filled circles) from the inset to Fig.~7~(c).}
\label{Fig:KL}
\end{figure}

\vspace{-2pt}\subsection{Calculation of lattice thermal conductivity}\vspace{-2pt}

The total scattering rate employed in the $\kappa$ calculations was a sum of the phonon-magnon scattering rate and a boundary scattering term, $\tau_{\rm tot}^{-1}=\tau_{\mathbf{q}}^{-1}+c/\ell_0$ (where $\ell_0=0.2$~mm was taken as a typical dimension of the single-crystal specimen transverse to the heat flow). We have ignored any $T$ dependence of the ME constant and possible scattering from defects and phonons, but these should be reasonable assumptions at low $T$, particularly to assess changes with magnetic field. The thermal conductivities were calculated~as,
\begin{align*}
\kappa_{\rm L,\perp}&=\frac{k_{\rm B}}{\left(2\pi\right)^3}\left(\frac{k_BT}{\hbar}\right)^2\int (c\cos\theta\sin\phi)^2 \tau_{\rm tot} \frac{x^4e^x}{(e^x-1)^2}\,{\rm d}^3q;\\
\kappa_{\rm L,\|}&=\frac{k_{\rm B}}{\left(2\pi\right)^3}\left(\frac{k_BT}{\hbar}\right)^2\int (c\cos\theta)^2\tau_{\rm tot} \frac{x^4e^x}{(e^x-1)^2}\,{\rm d}^3q,
\end{align*}
where $x=\hbar cq/k_{\rm B}T$. Figure~\ref{Fig:KL} shows the calculated $\kappa_{\rm L}$ (open squares and curves) at the four field values corresponding to those for the dispersions. The computed anisotropy was very small ($\kappa_{\rm L,\bot}-\kappa_{\rm L,\|}$ less than 1\% of $\kappa_{\rm L,\bot}$), so only a single curve is shown at each field.

For the purpose of assessing changes in $\kappa_{\rm L}$ with field, we ignored the shift in $\kappa$ toward larger values throughout the $T$ range seen in experiment [Fig.~7~(a) and Ref.~22], and fixed the magnetoelastic constant for the calculations so that the conductivities matched the experimental data at $T=5$~K. Also shown in Fig.~\ref{Fig:KL} is the thermal conductivity computed with only the boundary scattering term included ($\times$'s). As the phonon-magnon scattering declines with decreasing temperature, the computed $\kappa_{\rm L}$'s interpolate toward the boundary scattering curve, yielding a plateau or shoulder. This feature yields a peak in $\kappa_{\rm L}/T$ (inset, Fig.~\ref{Fig:KL}) qualitatively similar to that of the data from Gu \textit{et al.}~\cite{GuZhao18} (though the maximum occurs at somewhat higher $T$ in the calculations). With the exception of the 3~T data which matches experiment well, the computed $\kappa_{\rm L}$ overestimates the experimental $\kappa$.

Most notably, the calculated $\kappa_{\rm L}$ at low-$T$ decreases with increasing field, failing to reproduce the nonmonotonic behavior of experiment [Fig.~7~(a) and Ref.~22]. This trend in the calculations is consistent with the dispersions (Fig.~\ref{Fig:Dispersions}) which show increasing low-energy magnon spectral weight (from which phonons can scatter) as the field increases to 6~T. Though $\Delta_{\rm PG}$ increases significantly in going from $B=0$ to $B=3$~T [clearly seen in Fig.~S1\,(a,\,c)], the magnon spectral weight is enhanced at lower energy for $B=3$~T, in part due to the splitting of the pseudo-Goldstone mode into two separate modes along the $k_z$ direction
[Fig.~5 and Fig.~S1\,(b,\,d)], and especially due to the magnon dispersion at small angles $\theta$ (discussed further below in connection with calculations of $\kappa_{\rm m}$ and shown in Fig.~\ref{Fig:SmallTheta}). At 8~T the entire spectrum is gapped and the phonon-magnon scattering is suppressed.

\vspace{-2pt}\subsection{Calculation of magnon thermal conductivity}\vspace{-2pt}

The magnon velocities were computed from the gradient of the computed magnon dispersion function $E(\phi,\theta,k)$,
$$
\hbar\vec{v}_k=\overrightarrow{\nabla}_{{\kern-1pt}k{\kern.5pt}}E=\frac{\partial E}{\partial k}\ \hat{k}+\frac{1}{k}\frac{\partial E}{\partial \theta} \ \hat{\theta}+
\frac{1}{k\sin(\theta)}\frac{\partial E}{\partial \phi}\ \hat{\phi}.
$$
With the applied field direction along $\hat{z}$, the integral expressions for $\kappa_{\perp}$ ($\overrightarrow{\nabla}T$ along $\hat{x}$) and $\kappa_\|$ ($\overrightarrow{\nabla}T$ along $\hat{z}$) include scalar products $(\vec{v}_k\cdot\hat{x})^2$ and $(\vec{v}_k\cdot\hat{z})^2 $, respectively:
\begin{align*}
\hbar\vec{v}_k\cdot\hat{x}&=\sin\theta\cos\phi \frac{\partial E}{\partial k}+\frac{\cos\theta\cos\phi}{k} \frac{\partial E}{\partial \theta} -\frac{\sin\phi}{k\sin\theta} \frac{\partial E}{\partial \phi};\\
\hbar\vec{v}_k\cdot\hat{z}&=\cos\theta \frac{\partial E}{\partial k}-\frac{\sin\theta}{k} \frac{\partial E}{\partial \theta}.
\end{align*}
\noindent
The magnon thermal conductivities are given as,
\begin{align*}
\kappa_{\rm m,\perp}&=\frac{k_{\rm B}}{4\left(2\pi\right)^3}\int (\vec{v}_k\cdot\hat{x})^2 \tau_{\rm m} \frac{x^2e^x}{(e^x-1)^2}\,{\rm d}^3k;\\
\kappa_{\rm m,\|}&=\frac{k_{\rm B}}{4\left(2\pi\right)^3}\int (\vec{v}_k\cdot\hat{z})^2\tau_{\rm m} \frac{x^2e^x}{(e^x-1)^2}\,{\rm d}^3k,
\end{align*}
\noindent where $x=E/k_{\rm B}T$, $\tau_{\rm m}$ is the magnon relaxation time (taken to be a constant, \mbox{$\tau_{\rm m}=10^{-10}$~s}), and the factor of 1/4 multiplying the integrals arises from the smaller primitive cell of the fcc magnetic lattice compared to the crystallographic lattice.

The computed thermal conductivities are shown as functions of temperature and field in Fig.~\ref{Fig:Km}. We find \mbox{$\kappa_{\rm m,\perp}>\kappa_{\rm m,\|}$} at all fields for \mbox{$T\lesssim 5$~K} [Fig.~\ref{Fig:Km}~(a)]. The field dependence [Fig.~\ref{Fig:Km}~(c)] exhibits a maximum near $B=3$~T for $\kappa_{\rm m,\perp}$, quite similar to that observed experimentally, though the occurrence of a maximum at higher field for the computed $\kappa_{\rm m,\|}$ is inconsistent with experiment where a somewhat lower field scale for the maximum (1.3~T) is observed for this orientation.

The anisotropy and nonmonotonic behavior of $\kappa_{\rm m}$ at low $T$ arises primarily from features of the dispersion for momenta nearly parallel to the spin spiral axis (small $\theta$), where magnon states appear at ever lower energies with decreasing $\theta$ as the dispersion evolves toward its gapless character at $\mathbf{k}=\mathbf{q}_{\rm h}$ (Fig.~\ref{Fig:SmallTheta}). The nonmonotonic behavior of $\kappa_{\rm m}$ can be traced to the nonmonotonic variation in $\partial E/\partial\kern-1pt\phi$ with applied field (the amplitude of the oscillations that determine $\partial E/\partial\phi$ is largest for the $B=3$~T dispersions, highlighted by the vertical arrows in Fig.~\ref{Fig:SmallTheta}\,---\,a substantial contribution to $\kappa_{\rm m,\perp}$ from this term arises from the $(\vec{v}_k\cdot\hat{x})^2$ factor in the integral expression above.

\onecolumngrid

\begin{figure*}[h]
\centerline{\includegraphics[width=\linewidth]{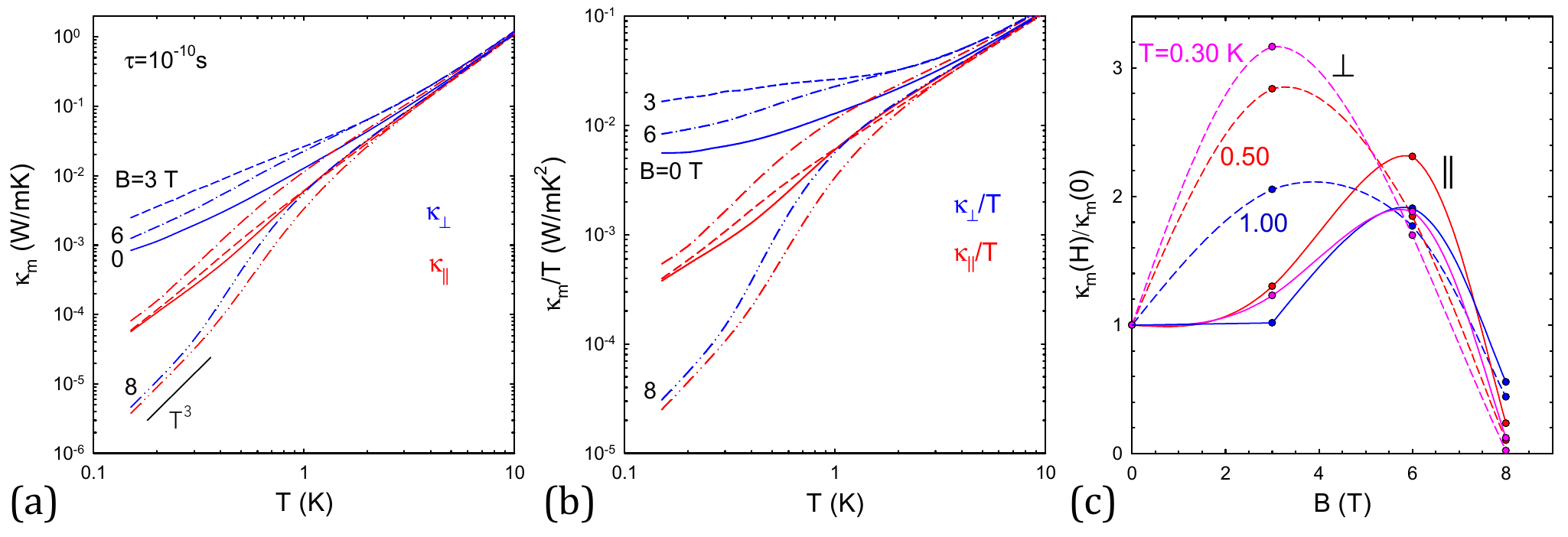}}
\caption{Computed (a)~$\kappa_{\rm m}$ and (b)~$\kappa_{\rm m}/T$ perpendicular ($\kappa_{\perp}$, blue curves) and parallel ($\kappa_{\|}$, red curves) to applied magnetic field for $B=0$, 3, 6, and 8~T. (c)~Normalized change in $\kappa_{\perp}$ and $\kappa_{\|}$ vs. magnetic field at $T=0.30$, 0.50, and 1.00~K.}
\label{Fig:Km}
\end{figure*}
\begin{figure*}[h]
\centerline{\includegraphics[width=\linewidth]{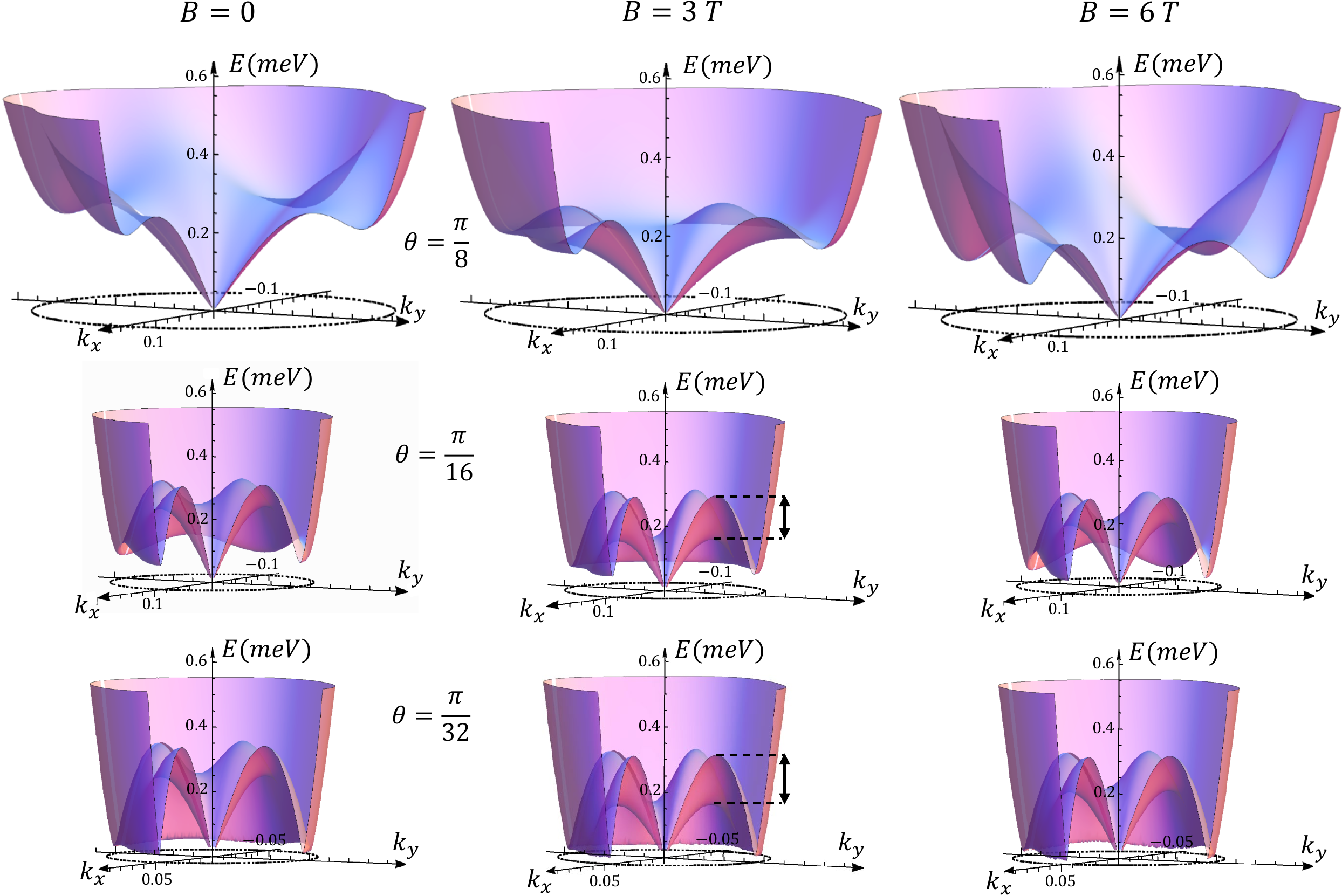}}
\caption{Magnon dispersions at $B=0, 3, 6$~T at angles (from top to bottom) $\theta=\pi/8, \pi/16,\pi/32$, projected into the $k_x-k_y$ plane (i.e. $k=\sqrt{k_x^2+k_y^2}/\sin\theta)$ to show the $\phi$ dependencies. The vertical arrows for $B=3$~T highlight the amplitude of the oscillations which determine
$\partial E/\partial\phi$.}
\label{Fig:SmallTheta}
\end{figure*}

\end{document}